# CHAPTER 5

## Galactic bulge-black hole co-evolution, feeding and feedback of AGNs

F. Combes, Observatoire de Paris, LERMA,
Collège de France, CNRS, PSL Univ., Sorbonne Univ., Paris, 75014, France

*Since the 1990s, we have known that there is a super-massive black hole in every galaxy, and that its mass is proportional to the mass of the bulge. To better understand how these black holes were formed, in symbiosis with their galaxies, we will look at their demography, the scaling relations between properties of black holes and host galaxies, and their evolution in a Hubble time. Observations at high angular resolution now allow us to enter the black hole sphere of influence, to see the molecular tori evoked in the AGN unification paradigm, and to understand the feeding processes of black holes. These are often accompanied by feedback processes, which moderate the formation of galaxies.*

## 1   Co-evolution: mass of black holes and bulges

### 1.1   How can we measure the mass of black holes?

If it is relatively easy to measure the mass of the black hole in our own Galaxy, the Milky Way (see chapter 2), this is because of its proximity, 8kpc. For galaxies belonging to the Local Group, it is already more difficult; the closest spiral, Andromeda, is distant by ~ 800kpc, so 100 times farther. And for the galaxies closest to our environment, in our local super-cluster Laniakea, the distance is around 10-80 Mpc, up to 10,000 times farther. It was only with the angular resolution of the Hubble Space Telescope (HST, 0.05-0.1''), launched in 1990, that it was possible to measure the velocity of stars and gas very close to the black hole, in its sphere of influence. The sphere of influence is defined as the area where the gravity of the black hole dominates that of the bulge, i.e., its radius is $r_s = G\ M_{BH}/\sigma_v^2$ = 10-50pc according to the mass of the galaxy. At a distance of 20Mpc, 0.1'' is equivalent to 10pc and we can solve the sphere of influence of the black hole. Later adaptive optics on ground-based telescopes made it possible to achieve comparable resolutions. To know the mass of the black hole, we must measure the velocity V of stars or gas at a certain radius R, and the mass will be proportional to the product $V^2\ R$,

One of the first relationships, linking the mass of the black hole and that of the bulge, was established by Magorrian et al. (1998) with 36 galaxies observed with the HST. They had already demonstrated a slightly different relationship for elliptical galaxies with or without cores. The core refers to the distribution of light in galaxies: if the density function ρ(R) is a power law (with slope 1-2), there is no core, but if the law ρ(R) is a plateau, we say that there is a core. The existence of cores could be due to the action of binary black holes, formed in the merger of galaxies that form elliptical galaxies. During a major merger of two galaxies each possessing a super-massive black hole, the dynamical friction causes the two galaxies to spiral towards each other, as well as the black holes, which begin to orbit fairly quickly in the center of the newly formed elliptical galaxy. Before merging, the pair of black holes must lose their relative angular momentum by dynamical friction on the stars of the spheroid: these stars gain energy, and are evacuated from the center, which creates a deficit of light in the center, and a core (Kormendy & Bender 2009, 2013). According to this scenario, galaxies with a core are the most massive, those with the least rotation, and appear to have been formed by merging galaxies without gas. The coreless elliptical galaxies were undoubtedly formed by merging gas-rich galaxies; the gas through dissipation fell towards the center and formed stars, creating an excess of light. These galaxies are less massive, and have more rotation.

Several tracers were used to determine the mass of the black hole: for the kinematic method, the velocity of stars, gas, and even masers in radio-astronomy (see Figure 1). The mass of the black hole M• is approximately 0.5 % of the mass of the bulge. What matters is only the mass of the bulge or spheroid, not the total mass, nor the mass of the disk for spiral galaxies, nor the mass of the dark matter halo (see Kormendy et al 2011).

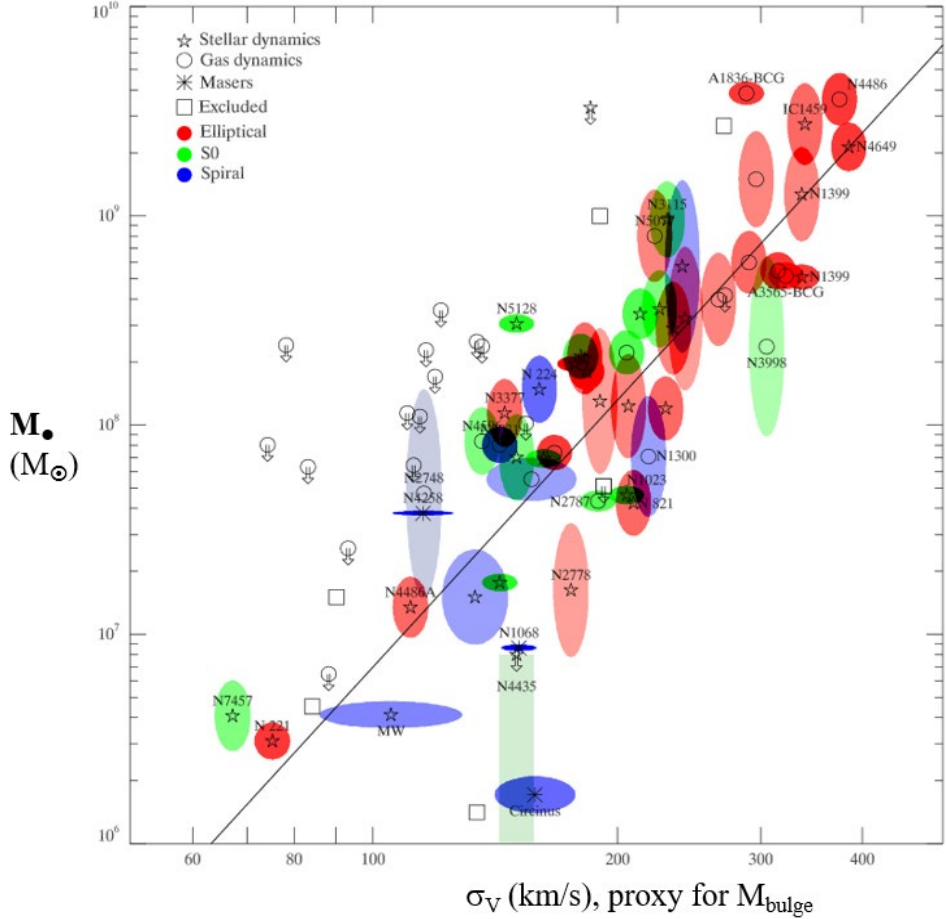

**Figure 1:** Relationship between the mass of the central black hole M• and the mass of the bulge, or the velocity dispersion ($\sigma_v$) in the center of the galaxy. Different symbols indicate the method of determination, and the color the categories of host galaxies. According to Gültekin et al. (2009).

The method based on the kinematics of stars is quite uncertain, because the stars in the center of elliptical galaxies are not only animated by rotation, but mainly by anisotropic velocity dispersion. The result then depends on the modeling, to determine the degree of anisotropy ($\sigma r$, $\sigma\theta$, $\sigma\phi$ dispersions), since the observations of the line of sight velocity are degenerated. The gas velocity does not suffer from this problem, but the gas is in general less massive, and easier to disturb, for example by the feedback of the active nucleus, hence the existence of outflows. The maser method is particularly precise and exact. A maser emission (the equivalent of a microwave laser) is a coherent emission of molecules (OH, $H_2O$, etc.), when they are all excited and emit simultaneously, in a stimulated manner. This can occur in regions of star formation, or the accretion disks around black holes, when the line of sight corresponds to a tangent to the disk seen edge-on, the place where molecules are crowded on the line of sight (see Figure 2). The emission is then very amplified, and it can be detected even at very high angular resolution, of around a milli arc-second by Very Long Baseline Interferometry (VLBI). A hundred galaxies have been observed in this way.

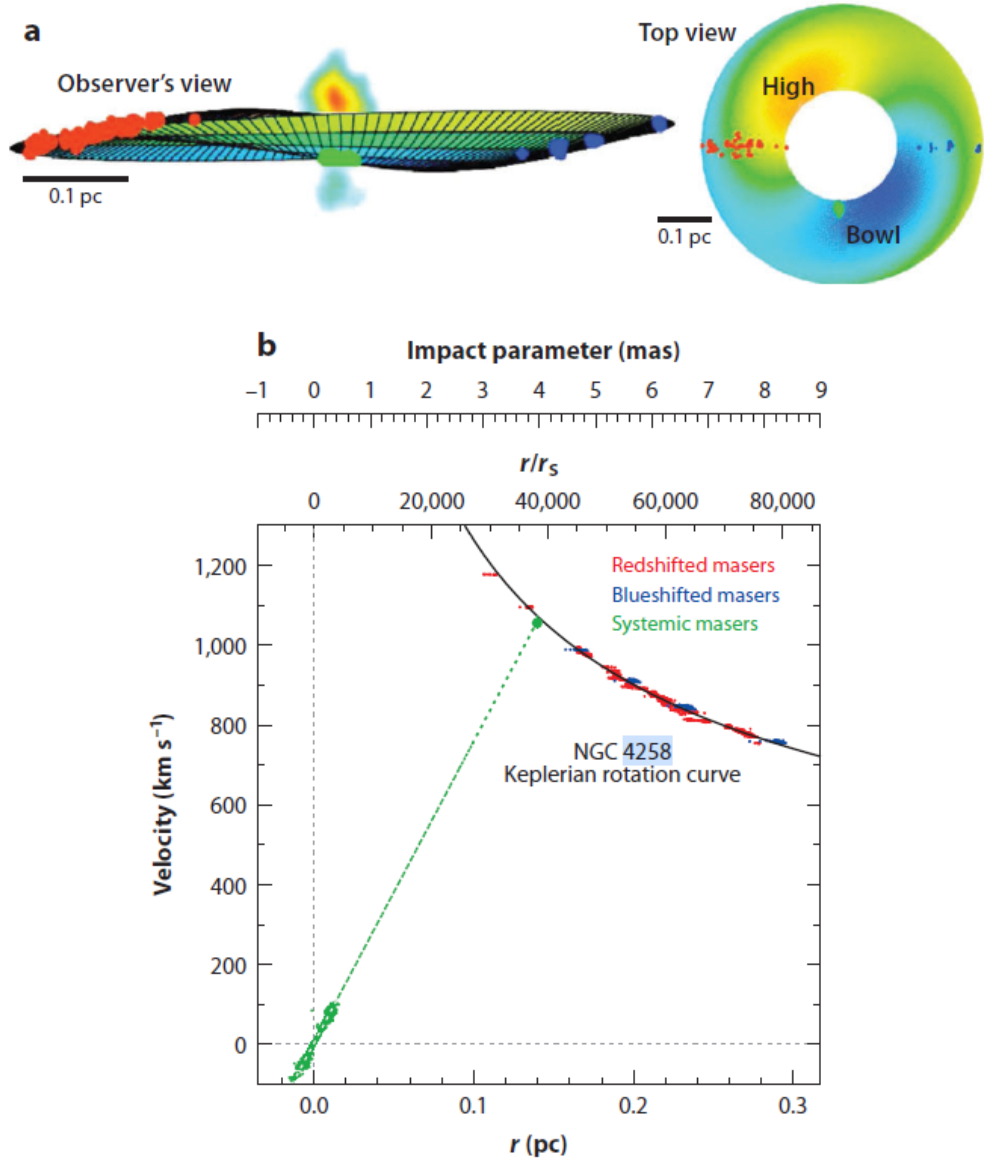

**Figure 2:** The galaxy NGC 4258 and its H$_2$O masers, discovered by Greenhill et al (1995): (a) Schema of the disk seen edge-on, showing a warp of the plane, with the radio jet perpendicular to it (after Moran 2008). In the face-on view on the right, we see the relief in colors (top in yellow, bottom in blue), and the positions of the maser emissions, according to their relative velocities (positive, zero and negative, red, green and blue points, respectively).

(b) the velocities are reported at the bottom on the rotation curve, which is very close to a Keplerian curve (expected around a point mass of 4 10$^7$ M$_\odot$), with some small uncertainties due to the de-projection. The scale is in parsec at the bottom, but also in milli-arc-second (mas) and Schwarzschild radius at the top (or radius of the horizon of the black hole r$_s$= 2 GM•/c$^2$). In green, the almost zero velocity is due to the masers drifting in the plane of the sky, we can then detect their proper motion, they move from right to left in 12 years over an area of ± 4° viewed from the center (Moran 2008).

The previous methods require angular resolution and can only be effective for nearby galaxies, whether the nucleus is active or not. For more distant galaxies, it is still possible to measure the mass of the black hole, if the nucleus is active, temporal and spectral resolution can replace spatial resolution. As long as the continuum emitted by the quasar or AGN varies, the variation propagates through the accretion disk into the Broad Line Region (BLR), very close to the black hole. The technique is called 'reverberation mapping'. The BLR responds up to radius R to brightness variations of the central continuum, which variably photoionizes it. Responses are seen in the lines after a time ~R/c shorter than the dynamic time tdyn = R/V (see Figure 3).

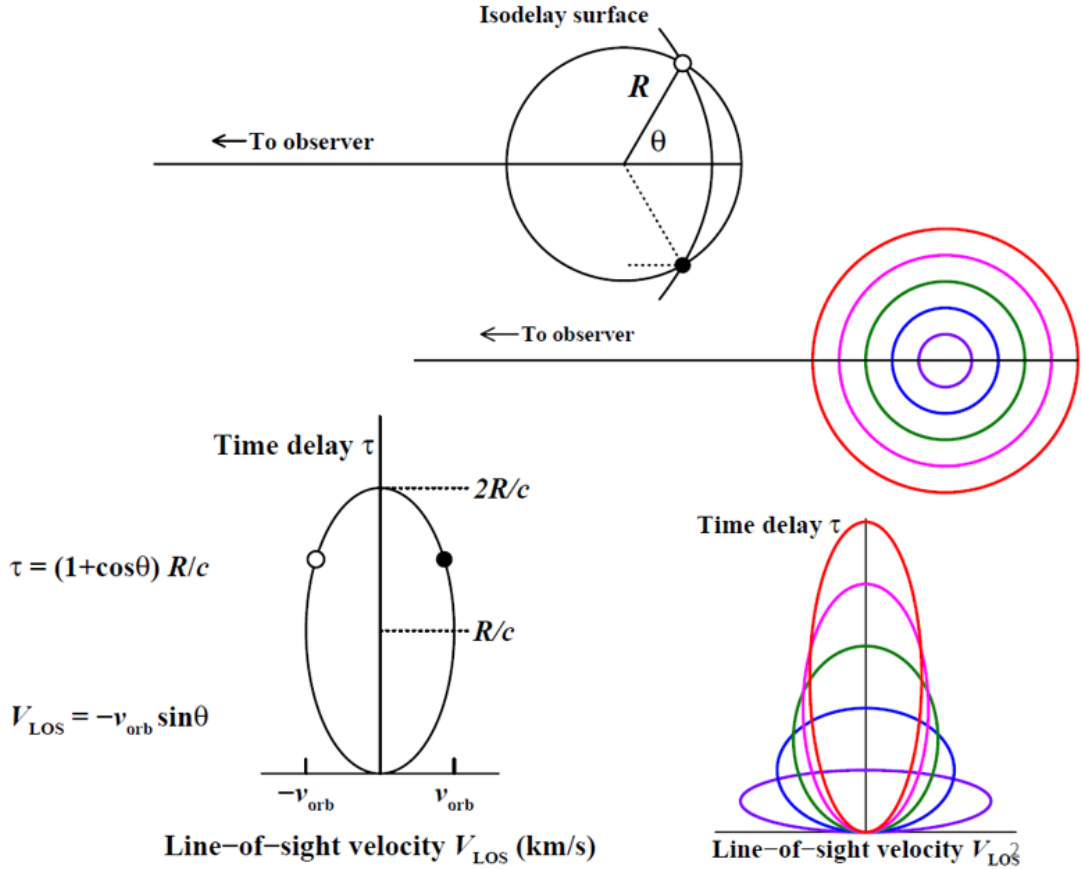

**Figure 3:** Principle of reverberation mapping. For a signal coming from radius R, the delay after a burst of the continuum coming from the origin is $\tau(R,\theta) = (1+\cos\theta) R/c$. Each line from the BLR (Hα, Hβ, etc.) has an optimum emission radius, symbolized by the various colors. Each angle θ corresponds to a Doppler velocity towards the observer $V_{LOS}$. We therefore observe the bottom-right diagram.

Blandford & McKee (1982) first used the term RM "reverberation mapping", for this method. A significant number of AGNs (about sixty) have been studied over time, thanks to numerous observations and long-term monitoring of variations, and the masses of their black holes have been precisely measured (e.g., Peterson et al 2004). Today, reverberation mapping is the most accurate "luxury" method for a distant AGN. It can be done on robotic telescopes. It is then used to calibrate the L-R relationship, Luminosity-Size of the BLR region, and thus deduce the masses of black holes from a much larger number of AGNs.

The relationship between quasar luminosity and BLR size is now well established by a power law of slope 2, i.e., $L \propto R^2$. This relationship can be explained if we assume that all bright AGNs have the same degree of ionization. The luminosity at a given frequency is proportional to the number of photons, which itself is proportional to the ionizing flux multiplied by the area $R^2$, and by the particle density or electron density $n_e$. This amounts to assuming that U and $n_e$ vary little from one AGN to another. Once measured the luminosity of the AGN (and therefore the size R of the BLR), and the width of the lines ΔV of the BLR region, we can apply the gravitational equilibrium, $M_{BH}= \Delta V^2 R /G$, within a projection factor of the velocity, because the orientation of the accretion disk is not known.

## 1.2   Bulge dispersion velocities

Soon enough, it was noticed that there was a much better relationship between the mass of the black hole and the central velocity dispersion of the bulge (Ferrarese & Merritt 2000). The velocity dispersion at the center is generally a function of the mass of the spheroid, a scaling relation called the Faber-Jackson relation for elliptical or early-type galaxies (Faber & Jackson 1976). It is the equivalent of the Tully-Fisher relation for spiral galaxies, where the visible mass is proportional to the 4$^{th}$ power of the rotational velocity, M (stars + gas) $\propto V^4$ (see McGaugh et al 2000).

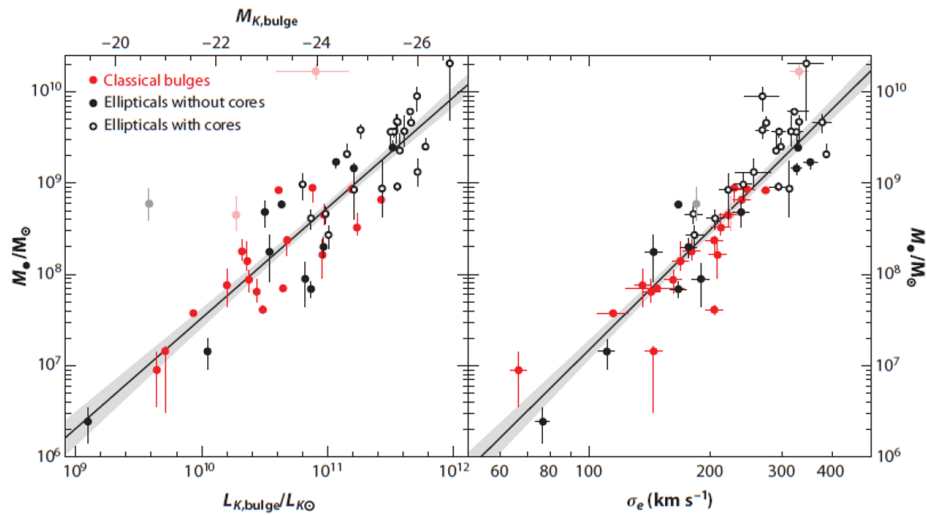

**Figure 4:** Comparison of the relation between $M_{BH}$ and total luminosity, with the relation $M_{BH}$ and velocity dispersion σe. A distinction is made between classical bulges of spirals (in red) and ellipticals in black. Empty circles are ellipticals with cores. For larger masses of black holes, $M_{BH}$ becomes almost independent of σe, at least for ellipticals with a core. According to Kormendy & Ho (2013).

In Figure 4, it is easy to see that the relation involving the velocity dispersion is less scattered. Note, however, that the effective velocity dispersion σe is defined as an average of the total kinetic energy, weighted at each radius by the corresponding luminosity (and therefore mass). More precisely, the square of the dispersion σe$^2$ is equal to the mean of $V^2(r)+\sigma^2(r)$, weighted by the luminosity at radius r. In the case where the mass of the black hole is determined by the kinematics of stars or gas, the two variables, black hole mass and dispersion, are not entirely independent, which may partly explain the weaker scatter of the relationship.

In addition, we can see in Figure 4 that there is a small distinction between ellipticals with or without a core. This is already found in the Faber-Jackson relation: if the coreless ellipticals do verify the law $L \propto \sigma_v^4$, the velocity dispersion is much lower for galaxies with cores, and the relation becomes rather $L \propto \sigma_v^8$. The interpretation is that the galaxy mergers that form the most massive ellipticals result in the merger of black holes, which in their coalescence heat and evacuate matter from the center, reducing the velocity dispersion (Kormendy & Bender 2013). The mass of the bulge then becomes independent of sigma, and the $M_{BH}$ – sigma relation saturates at large masses. Elliptical galaxies with cores have box shapes, and do not spin, while galaxies without a core have disk shapes and more rotation. These observations are fully consistent with the essential role of black hole mergers in core formation.

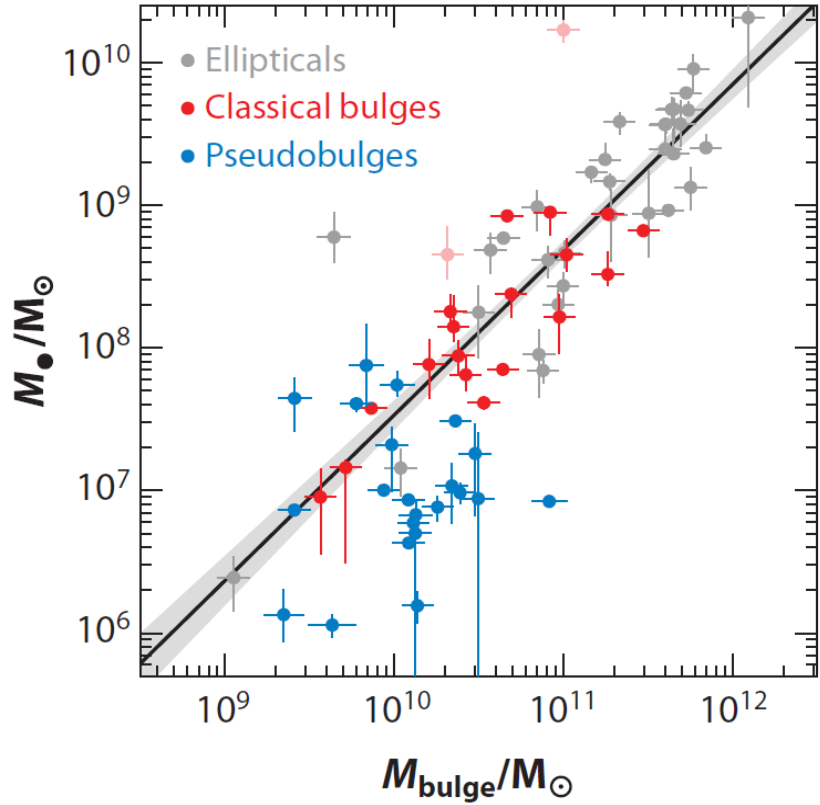

**Figure 5:** The proportionality between mass of the black hole and mass of the bulge is less scattered for massive spheroids, early-type galaxies. For galaxies of lower mass, the relation is with "classical" bulges and not pseudo-bulges of barred spiral galaxies, and since it is difficult to separate them, the relation is much more scattered. According to Kormendy & Ho (2013).

Another distinction can be made between classical bulges, which are believed to be the result of minor mergers of galaxies and pseudo-bulges, resulting from the secular evolution of spiral disks. Figure 5 shows that the relationship between mass of the black hole and mass of the bulge mainly concerns classical bulges. However, the relationship becomes very scattered at small mass. This could come from a determination problem, because for small masses, the black hole sphere of influence is very small, and requires very high spatial resolution. Also, the activity of nuclei is weak, and it can be confused with other activities, such as star formation. The mass of nuclear disks, whether stellar or gaseous, becomes comparable to the mass of the black hole.

For spiral galaxies, other black hole mass indicators have been proposed, such as the pitch angle of the spiral arms (Davis et al. 2014), or the Sérsic index (Mutlu-Pakdil et al. 2016). The relationships are empirical, but could be justified by the indirect correlation between the tightening of the spiral arms and the stability of the disk, which is increased by a massive bulge.

It is even more difficult to measure the mass of a possible black hole, of intermediate mass, in dwarf galaxies. However, a certain number of dwarfs have an active nucleus, detected thanks to its x-ray, optical/infrared or radio emission (see the review by Reines & Comastri 2016). It is difficult to distinguish AGNs in dwarfs, however, as the diagnostics can also apply to a region of starburst (Hainline et al 2016). The relationship between mass of the black hole and mass of the bulge seems to continue even at low masses (Baldassare et al 2016).

## 1.3  Special cases and exceptions: obese black holes

There are exceptions to the general law, for example a number of systems have black holes that are too light for their velocity dispersion. The vast majority of these are interacting galaxies that will merge into one. The velocity dispersion has already increased, but the black holes have not yet merged, and the final system will reverse the relationship, after less than a billion years.

There is also a whole category of objects, where the black hole is obese, compared to the host galaxy. These galaxies are in clusters, and they are often the brightest galaxy in the cluster. These galaxies are like cannibals in the center of the cluster, they merge with lots of little companions, and most notably, swallow hot gas before it has even had time to form stars. This process could cause the black hole to grow faster than the bulge. One of the examples of obese black-hole galaxies is NGC 1277 in the Perseus cluster. It is a lenticular galaxy, possessing a rotating disk, and a massive bulge. Van den Bosch et al (2012) measured a giant black hole mass of $1.7 \; 10^{10} \; M_\odot$, or more than half of the mass of the bulge, instead of the standard ratio of 0.5%. This large value has been questioned by Emsellem (2013), reducing the mass of the black hole by a factor of 4, which was evaluated using stellar kinematics, and proposing a higher bulge mass, and perhaps a velocity disturbance due to a bar on the line of sight. The mass of the black hole was, however, confirmed with molecular gas kinematics (Scharwächter et al 2016).

Other examples are found in the Perseus cluster, and also in the core of other clusters. If the ratio of the mass of the black hole to the mass of the bulge is too large, it would appear to be due to a deficit of star formation in the bulge, rather than an excess of mass of the black hole. In clusters, star formation can be easily stopped, as the very hot gas from the cluster acts as a wind that sweeps away the gas from galaxies. Likewise, galaxies can be suffocated if they can no longer replenish themselves with cold gas, as the filaments of matter and gas are heated and destroyed at the entrance of the cluster. Observations support the scenario, because the stellar population in these galaxies is very old. These could be "relic" galaxies, where all of the stars were formed at the beginning of the Universe, and which could then have acquired gas to fuel their black hole, but without forming new stars (see Trujillo et al 2014).

The difficulty in determining the mass of black holes using stellar dynamics is well illustrated by the example of our nearest spiral galaxy: Andromeda or M31. Its proximity makes it a very favorable object, because with the HST, we have excellent angular resolution. At a distance of 780 kpc, 0.1 arcsec represents 0.37 parsec, and the kinematics can be observed well within the sphere of influence (r~20 pc) of the black hole, which is approximately $10^8$ M$_\odot$. However, there are difficulties which arise precisely because of this proximity. We can see that the dynamics of the stars in the center are not in equilibrium. There is an eccentric nuclear disk of stars, 2 pc in diameter. The black hole is not in the center of the disk. This asymmetry is probably generated by a density wave, which rotates with a high angular velocity.

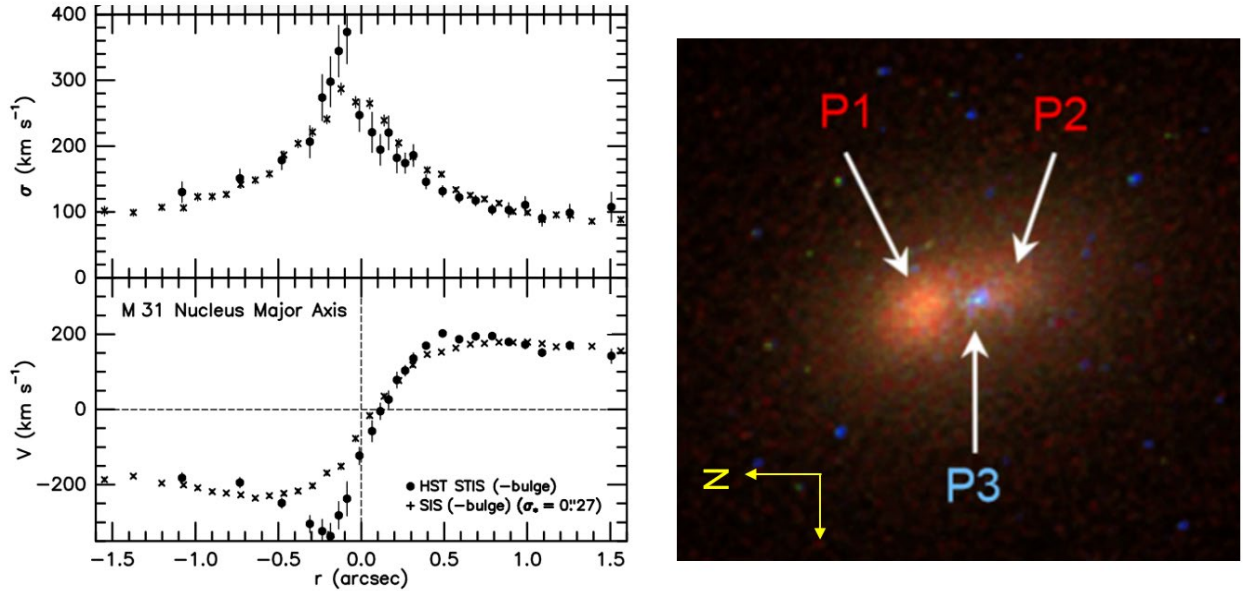

**Figure 6:** The black hole in the center of our neighboring galaxy M31: On the left, the stellar kinematics, obtained with the HST by Bender et al (2005), with the dispersion at the top, reaching a maximum of 400 km/s, and the rotation along the major axis at the bottom. On the right, the photometry of the HST, showing the three light components of the eccentric nuclear stellar disk, according to Lauer et al (2012). The black hole is in P2, which is not the brightest point.

Figure 6 shows the arrangement of the eccentric nuclear stellar disc. The brightest point on the disc is P1. The black hole, detected by the maximum of the velocity dispersion, is in P2. The separation between P1 and P2 is 0.5'' = 1.8 pc. The rotational velocity is V=200 km/s, with an orbital period of 50,000 years. P3 is a very young blue star cluster, no more than 200 million years old. The kinematic major axis of the velocity field does not correspond to the photometric major axis, symbolized by the segment P1-P2. The tilt of the stellar disk is uncertain, but would rather be 55°, quite different from that of the M31 galaxy, which is very tilted (i=77°). The system should not be stable, and our observation of this state would be highly unlikely. For example, star clusters should merge, unless there is a dynamical mechanism to lengthen the lifespan of such a distribution.

This mechanism is probably a wave of symmetry m=1 (Tremaine, 1995); these are the most frequent waves to develop in a Keplerian disc, i.e., dominated by a point mass. N-body simulations have shown that such a wave is compatible with the observations (Bacon et al 2001). In the simulations, the nuclear disk of M31 is subject to a natural mode, with a very slow rotational pattern speed (3 km/s/pc, period of 2 million years) which can be maintained for over a thousand dynamical times, or periods of rotation. The resulting morphology and kinematics can reproduce the photometry of the nuclear disk observed in M31 and the average stellar velocity, with its asymmetries. They require a central mass concentration (the black hole) and also a cold stellar disk of 20 to 40% of the mass of the black hole. Such a slow mode could be excited when interstellar clouds from the outermost gaseous disk infiltrate towards the center. Nuclear disks formed from accumulated gas are possible candidates for the precursors of these types of structures and may be common in central regions of galaxies. Peiris & Tremaine (2003) simulated a massless, non-self-gravitating nuclear disk. They tried several inclinations and alignments with the main disk of the Andromeda galaxy. Good agreement with observations is only obtained when the stellar disk is not aligned.

## 1.4 High redshift black holes

Does the relationship between the masses of the black hole and bulge persist during evolution, and when did it become established? A large number of quasars have been observed at large z values, even greater than 6, and the dynamics of the host galaxy could be determined by the CO and [CII] lines, which are redshifted into the millimeter range (observable with ALMA or NOEMA interferometers). The mass of their black hole is determined by optical observation of the spectrum of their BLR region, and the relationship between brightness and size of the BLR region, described earlier in this chapter. For the most part, the mass of high z black holes is greater than expected from their dynamical mass, and from the relationship obtained locally (see Figure 7, Venemans et al, 2016). This surprising result could come from the great uncertainties when estimating the dynamical mass. It is not possible to distinguish the bulge from the disk, for example, and the mass of the black hole is compared to the total dynamical mass of the host. Therefore, the mass of the bulge could only be less, which would increase the distance from the local relation. Another uncertainty is that of the host galaxy morphology. The inclination of the rotating molecular disk is not well known and the result is only valid statistically. Estimating the mass of the black hole, using the brightness-size relationship of the BLR region, also has a factor of uncertainty due to unknown inclination. In most systems, it is assumed that the [CII] or CO lines are centered on the systemic velocity of the host galaxy, and can be reliably used to calculate dynamical mass. However, the broad MgII optical emission lines are systematically shifted towards the blue. The average shift is 500 km/s, but can go up to 1700 km/s. It is interpreted as an outflow due to feedback from the central AGN; the symmetrical red-shifted region is too obscured by dust to be detected. However, even if the dynamical mass is not well estimated, it is possible to have a lower limit with the gas mass, from the two lines CO and [CII] and also from the emission of the dust. In these large-z systems, the gas fraction is often greater than 50%, and the dynamical mass can only be underestimated by a factor of 2 at most. The masses of the black holes are then 3-4 times higher than expected according to the local relation.

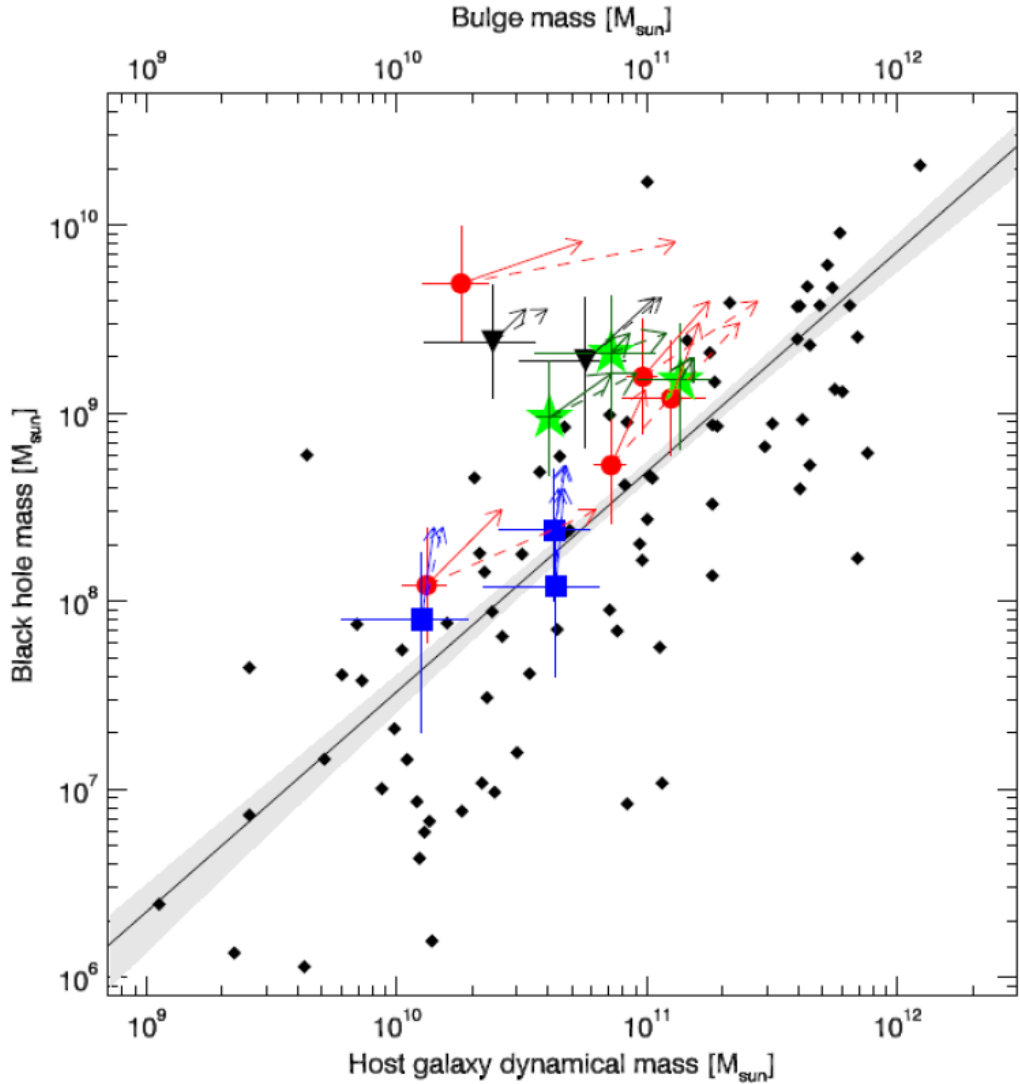

**Figure 7:** The mass of the black hole as a function of the dynamical mass of distant galaxies (z = 6) hosting quasars, and for local galaxies, as a function of the mass of the bulge (top scale). The black diamonds represent local galaxies (Kormendy and Ho, 2013). Their relationship is underlined by the shaded area. Large colored symbols are high z quasars, which overwhelmingly have excess black hole masses. According to the accretion rate of matter deduced from the luminosity of the quasar, and taking into account the observed star formation rate, it is possible to extrapolate the trajectory of the points (arrows) over the next 50 million years. According to Venemans et al (2016),

## 2   Interpretations: cause or effect?

As presented at the beginning of this chapter, there is locally a strong relationship between masses of black holes and masses of bulges. There is also a simultaneity of evolution between star formation, or the mass growth of bulges, and the activity of nuclei, or the mass growth of black holes. These two combined relationships suggest a symbiotic formation of galaxies with their black holes.

## 2.1 Evolution of star formation and the accretion of black holes

The distribution of active star-forming galaxies showed a very strong evolution with the redshift. The majority of the stars we see in local galaxies were formed in the first half of the Universe's life. A parallel evolution is observed for the activity of the nuclei. While bright quasars are quite rare locally, they were much more common at the beginning of the Universe. The activity of nuclei is a tracer of mass accretion by super-massive black holes at the center of each galaxy, and we can use the continuity argument (Soltan 1982) to associate the growth rate of black holes to the number and total luminosity of quasars over time. For this we use the average radiation efficiency of an AGN: any mass m swallowed by a black hole, will radiate $\varepsilon = 10\%$ of its mass energy $mc^2$ and thus contribute to the brightness of the quasar. There is also an efficiency term, because some nuclei may have reduced activity, for the same rate of accretion (for example Advection Dominated Accretion Flow (ADAF), see Chapter 3). For this black hole mass growth rate, we mean the overall rate, and the merging of galaxies leading to the merging of black holes, without an increase in their combined mass does not count.

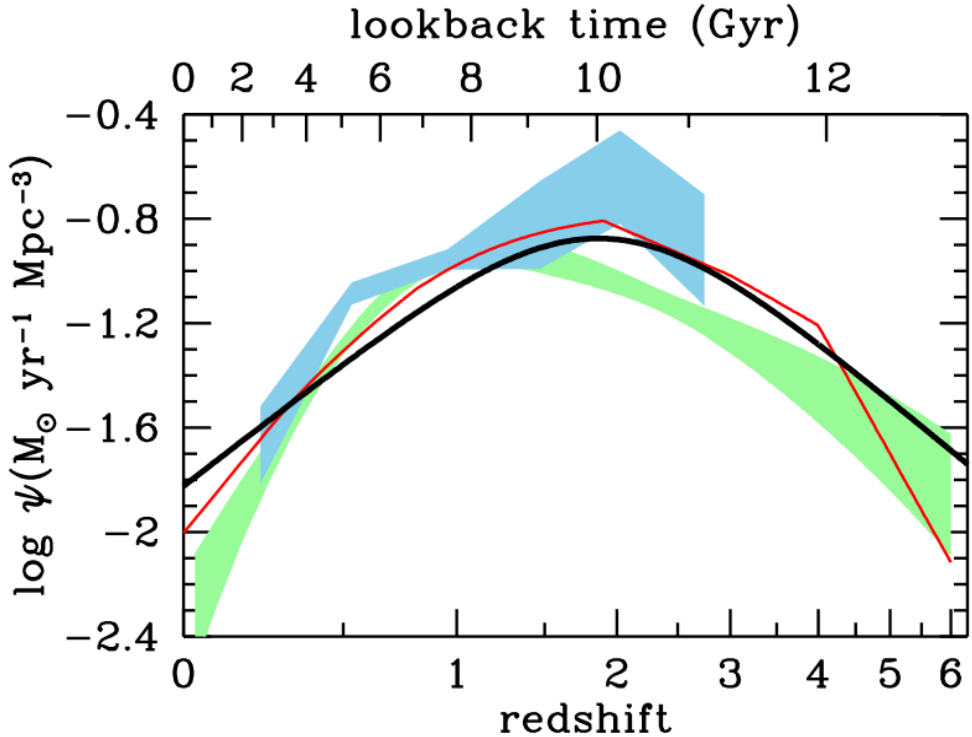

**Figure 8:** The cosmic evolution of star formation density is the thick black curve. The density increases to a maximum of z = 2, then decreases by a factor of ~ 10 until today. The activity of quasars and AGNs follows a parallel evolution: the red curve and the green area are two estimates using the x-ray indicator; the blue area is derived from the infrared domain. The rate of mass accretion on black holes has been multiplied by 3300 to enter the same scale as the density of star formation. According to Madau & Dickinson (2014).

How can we monitor the evolution of black holes over time? The main source of information is studying the active nuclei as a function of redshift, to establish the all-z luminosity function, and to obtain the mass of black holes from the spectrum of the BLR, and the brightness-size relation of the BLR. If the mass accretion rate is dmacc/dt, the radiation efficiency is ε, the luminosity of the AGN is L = ε dmacc/dt c². In this accretion, the mass of the black hole increases by dM•/dt = (1-ε) dmacc/dt, and the link between the increase in mass of black holes and the brightness of the AGN is dM•/dt = (1-ε)/ε L/c².

Normally, this inventory would account for the mass function of black holes today at z = 0, based on the relation M•-M$_{bulge}$ and the bulge mass function, regardless of nucleus activity. Yet the comparison does not match, there is twice the mass of black holes today compared to the integrated luminosity of all AGNs, and it must be assumed that the mass accretion by black holes is not always radiatively efficient (Fabian & Rees 1995). Or it is possible that much of the radiation from AGNs is dust obscured and has escaped observation.

Another more global method is to observe the accumulated background radiation, especially in x-rays, by AGNs during evolution.

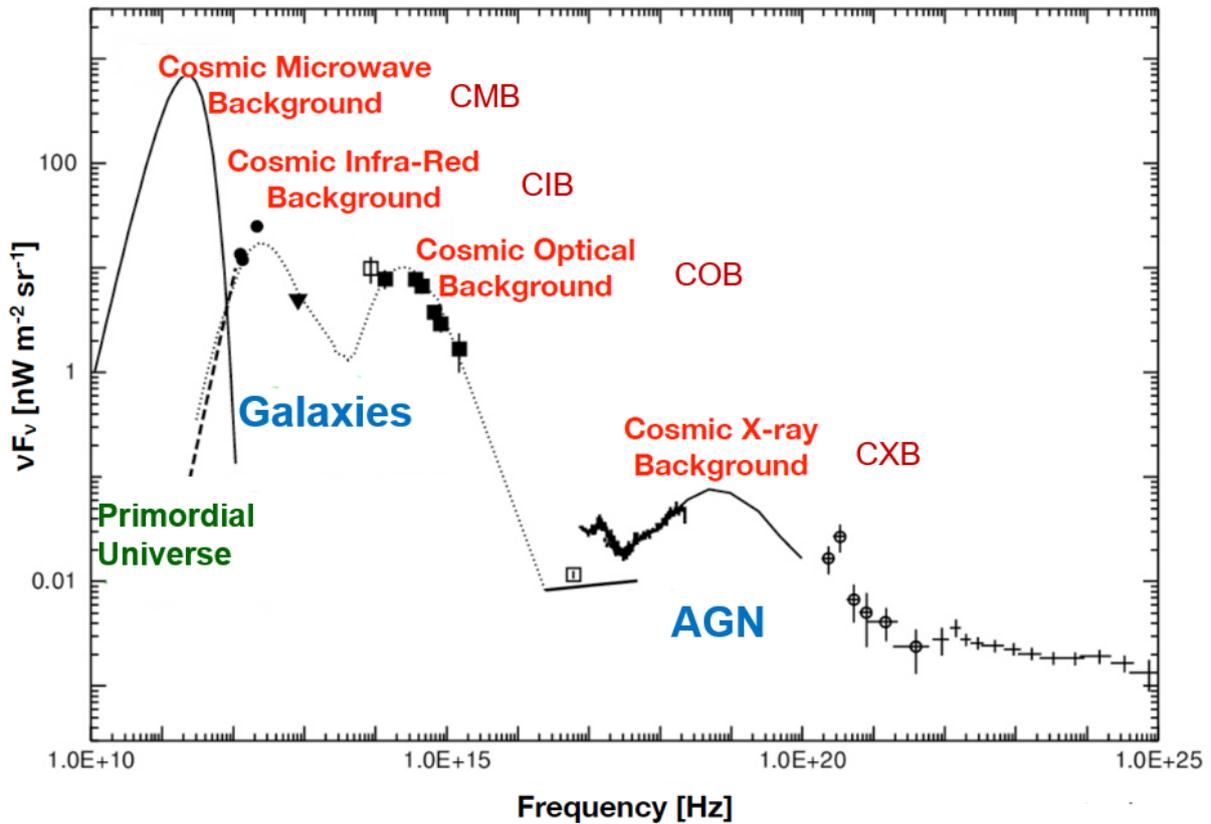

**Figure 9:** The spectral density of cosmic energy, from radio waves to the highest gamma frequencies. Representation in νFν weights the radiation with its energy, and with the log-log scale a horizontal line corresponds to radiation of equal power per decade of energy. According to Hasinger (2000).

Figure 9 shows the entire spectrum of cosmic background radiation. The energy is dominated by microwave background radiation coming from the Big-Bang, then the optical and infrared background radiation come from the radiation of stars in the galaxies, and from dust heated by young stars and also by AGNs. X-ray radiation is dominated by AGNs, but the spectrum of harder radiation indicates that much of this radiation is obscured by dust, so it is re-radiated in the far infrared or sub-millimeter range, at lower frequency. X-rays can detect obscured AGNs, but up to a column density of $10^{24}$cm$^{-2}$. Beyond that, for higher column densities, even the X-rays do not get out, these are so-called "Compton-thick" sources. To make an inventory of all AGNs, all these corrections must be made.

## 2.2 Downsizing, formation of AGNs coupled with the formation of stars

The evolution of the luminosity functions of AGNs reveals an anti-hierarchical effect called "downsizing" (Cowie et al 1996). This effect has also appeared for star formation: the most active galaxies today are the less massive galaxies, the late-type spirals and the irregulars. The brightness of the most active galaxies continues to decrease over time. However, in the standard cosmological model, which is hierarchical, the objects that form first are the smallest, they then merge to form the largest, but these are dark matter halos. As for black holes, the most massive ones appear earlier in the Universe, and are now silent, while activity continues in Seyfert galaxies, in general of spiral types.

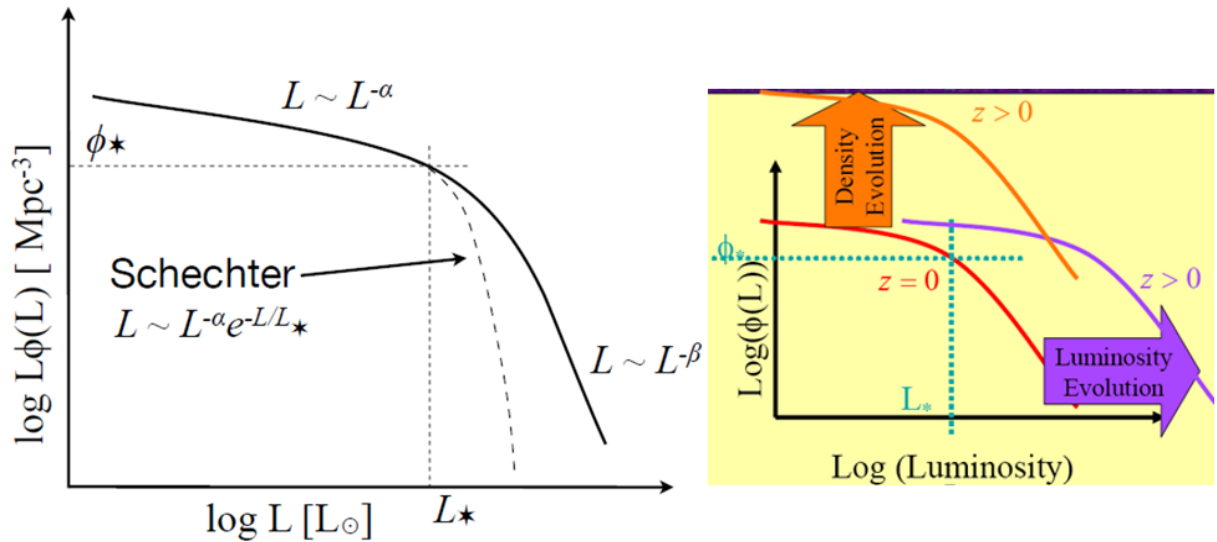

**Figure 10:** Left: The AGN luminosity function is a double power law, $\alpha$ for low luminosities (L<L*) and $\beta$ for higher values. It can be written as $\Phi(L) = \Phi(L^*) / [(L/L^*)^\alpha + (L/L^*)^\beta]$, with $\alpha < 1$ and $\beta > 1$. It is very different from the luminosity function of galaxies, for the brightest objects. Right: the evolution of the luminosity function with the redshift breaks down into an evolution of the luminosity (L* increases), and also the density of AGNs ($\Phi^*$ increases).

The luminosity function of AGNs can be described by a double power law, with two slopes $\alpha$ and $\beta$ (see Figure 10). It is different from the luminosity function of galaxies, which is a Schechter law, first a power law of slope $\alpha$ at low luminosity (L<L*), then an exponential law, in exp (-L/L*). The brightness function of AGNs varies significantly with redshift, both in brightness and density (see Fig 10). The two effects cannot be separated. On the other hand, the two slopes $\alpha$ and $\beta$ also change with the redshift, and that is what causes the downsizing. Observations show that AGNs of higher masses and luminosities form earlier (anti-hierarchical formation), this implies coupled evolutions of luminosity and density, and an evolution of slopes $\alpha$ and $\beta$ with z. A model reproducing the X-ray, optical and infrared observations, therefore in bolometric luminosity of AGNs, clearly shows the anti-hierarchy and the variation of slopes α and β (Figure 11). The AGN activity maximum is well reproduced at redshift equal to 2, as in Figure 8.

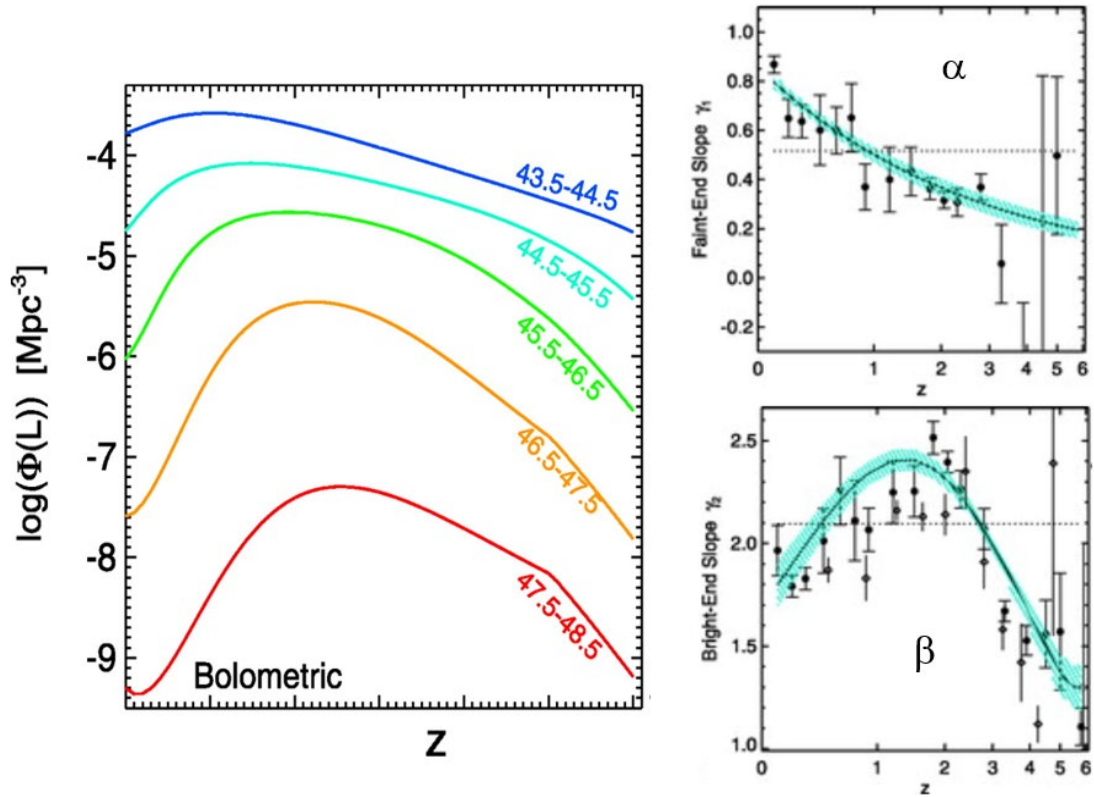

**Figure 11:** On the left, the various curves correspond to a given bolometric luminosity, in erg/s (only the log is indicated). For the brightest AGNs, the maximum occurs at greater redshift than for weaker AGNs. This is the anti-hierarchical evolution (downsizing). In this model, it is necessary that the slopes of the brightness function of the quasars, α and β, evolve as a function of z, as shown on the right. According to Hopkins et al (2007).

Regarding the mass of black holes, the evolution is well reproduced, assuming an average radiation efficiency ε = 0.04-0.16, with an accretion rate such that the active phase of AGNs occurs with a luminosity equal to 0.1-1.7 of the Eddington luminosity. The lifespan of these active phases varies between $1.5\ 10^8$ years for large black hole masses (greater than $10^8$ M$_\odot$), and $4.5\ 10^8$ years for the smallest (Marconi et al 2004). For the anti-hierarchy, massive black holes are formed in the deepest potential wells, which form first. The lightest black holes form in shallow wells. They are formed later in the Universe, and moreover they are more sensitive to the feedback phenomenon. They therefore take much longer to grow.

In the 1990s, during the first far infrared sky survey, using the IRAS satellite, a new phase in the evolution of galaxies was discovered, a huge burst of star formation: these objects are called "starburst galaxies". They had not been suspected before, as their immense star-forming activity is almost entirely obscured by dust, and emerges in far infrared (Sanders & Mirabel 1996). They are ultra-luminous infra-red galaxies, or ULIRGs ($L_{IR} > 10^{12} L_\odot$). At these very strong luminosities, almost all objects are the result of interacting or coalescing galaxies. In the interaction, the tidal torques were able to drive enormous amounts of interstellar gas towards the center of the galaxies, and produce bursts of star formation. In these very compact areas, the dust is concentrated, and re-emits in the far infrared the light from buried young stars. The gas is not entirely consumed by star formation, and can also fuel central black holes. The ULIRG phase could thus be a precursor of the quasar activity phase. The correspondence between the luminosities of extreme starbursts and quasars is compatible with this hypothesis. The activity phase of the quasar is shorter as the feedback is more violent, which corresponds to the observations.

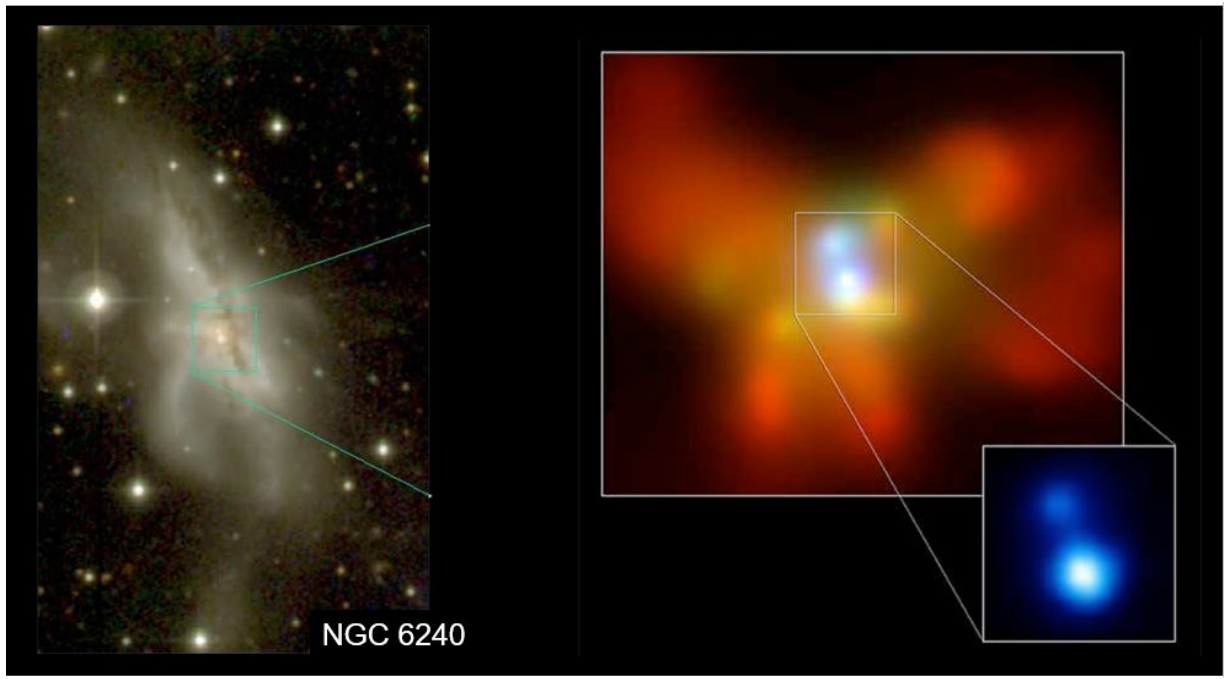

**Figure 12:** The advanced coalescing two-galaxy system, NGC6240 (HST image, left), is an ULIRG, a starburst galaxy. It is the prototype of starbursts buried in dust and molecular gas. The near infrared emission of the $H_2$ molecule could be detected, and at high angular resolution, two nuclei appear, emitting strongly in X-rays, indicating that the merger is not complete. The powerful and hard X-rays confirm that both nuclei are AGNs. Two iron atom recombination lines (Kα), in fluorescence near the two black holes, were also detected (Komossa et al 2003).

However, there is still some degeneracies when interpreting the luminosity function of quasars. Two models reproduce the observed luminosity function: either there are many short-lived quasars, or there are rare long-lived quasars. A constraint on the parameters is already given by the mass function of black holes established locally. Mass accretion must be relatively shared among all galaxies during evolution, since black holes with extreme masses (greater than $10^{10} M_\odot$) are not often observed. However, there are several free parameters, not only the radiation efficiency ε, but also the Eddington rate: i.e., does the growth of black holes mainly take place near the maximum set by the radiation pressure (see chapter 1)? The luminosity function of AGNs in two-slope power law α and β has been interpreted on the one hand (α) by more regular growth away from the Eddington rate, and (β) more intermittent growth at the Eddington rate, during galaxy mergers. These would be the same galaxies seen in these two emission regimes, but in different phases of their evolution. Since both low and high luminosity AGNs come from the same population, a wide range of brightness comes from a small domain of the mass of galaxies and black halos. The consequence is that the degree of agglomeration of quasars, in groups or clusters, should not depend on their luminosity. An attempt at clarification could be made by comparing the agglomeration rate of AGNs with that of dark matter distribution. If quasars are activated by merging galaxies, the maximum attainable luminosity is correlated with the mass of the halo, but not the instantaneous luminosity. The detailed study of the simulated light curves, the degree of correlation with the mass of the halo, and the comparison with the agglomeration of black halos and galaxies showed that the agglomeration rate of AGNs had very little dependence on luminosity, which is compatible with the predictions of the model (Lidz et al 2006). Quasars cluster less than elliptical (red) galaxies but more than spiral (blue) galaxies. They form an intermediate population.

## 2.3  Feedback: how can the black hole influence the bulge?

Many models propose to explain the excellent correlation between masses of black holes and masses of bulges through the AGN feedback phenomenon: as soon as interstellar gas arrives towards the center, it can both form stars and fuel a black hole, but as soon as the latter is active, its energy will heat the gas or eject it mechanically, preventing any subsequent star formation and therefore limiting the mass of the bulge. Yet the radius of the black hole sphere of influence $R_{infl} = GM_\bullet/<V^2>$, defined as a function of the average velocity or dispersion velocity $<V^2>$ of the bulge at radius R, is extremely small; approximately 20 pc, much smaller than the radius of the bulge (on the order of kpc). More precisely, the observed mass correlation is $M_\bullet \sim 0.002\ M_{bulge}$, with the mass of the bulge equal to $M_{bulge} \sim 5 <V^2> R_{bulge}/G$, i.e., $R_{infl} \sim 10^{-2}\ R_{bulge}$. The volume of influence of the black hole is therefore $10^{-6}$ times the volume of the bulge. Under these conditions, it is difficult to consider a gravitational exchange of information.

However, when you calculate the energy returned to the rest of the galaxy as the black hole grows, accreting at a dmacc/dt rate and with an average efficiency of $\varepsilon \sim 0.1$, and an AGN luminosity of $L = \varepsilon\ dmacc/dt\ c^2$, the growth of the corresponding black hole mass is $dM_\bullet/dt = (1-\varepsilon)\ dmacc/dt$. The energy expended during the growth of the black hole is therefore $E_c = \varepsilon/(1-\varepsilon)\ M_\bullet\ c^2$. This can be compared with the gravitational energy of the bulge $E_{bulge} \sim M_{bulge} <V^2>$. It appears that the growth energies of the black hole and the bulge are in the ratio $E_c/E_{bulge} = \varepsilon/(1-\varepsilon)\ (M_\bullet/M_{bulge})\ c^2/<V^2> \sim 400$!

The active nucleus deploys enough energy to destroy the bulge. The whole problem is therefore to know what fraction of this energy will be radiated outwards without affecting the galaxy, and what fraction will be coupled with the disk or the bulge, thus justifying an interaction between the two, and a possible moderation. We will now identify the various possible models of coupling between AGN and galaxies, depending on the brightness of the AGN.

## 2.4  Feedback: radiative mode and quasar wind, or mechanical mode of the radio jet

One of the fairly obvious modes of interaction between an AGN and the galaxy is through the wind evaporating from the accretion disk, when the luminosity of the quasar is very high, near the Eddington limit. Note that Eddington luminosity is obtained when the radiation pressure compensates for gravity, the gas can no longer flow freely and feed the black hole. This Eddington limit is $L_{Edd} = 3.3 \times 10^{13} L_\odot\ (M/10^9 M_\odot)$, for a black hole of $10^9 M_\odot$ for example. A black hole, with matter around it distributed spherically, will not be able to radiate beyond it.

This is possible, however, if the material is distributed on a disk. The accretion rate corresponding to this limit is 20M☉/year (M/$10^9$M☉).

In the case of very bright quasars, it is therefore common for a wind of matter to rise above the disk (see Figure 13). For all the others, the possible mechanism would rather be mechanical, through driving the jet. Note, however, that radio jets are only observed in 10% of all AGNs.

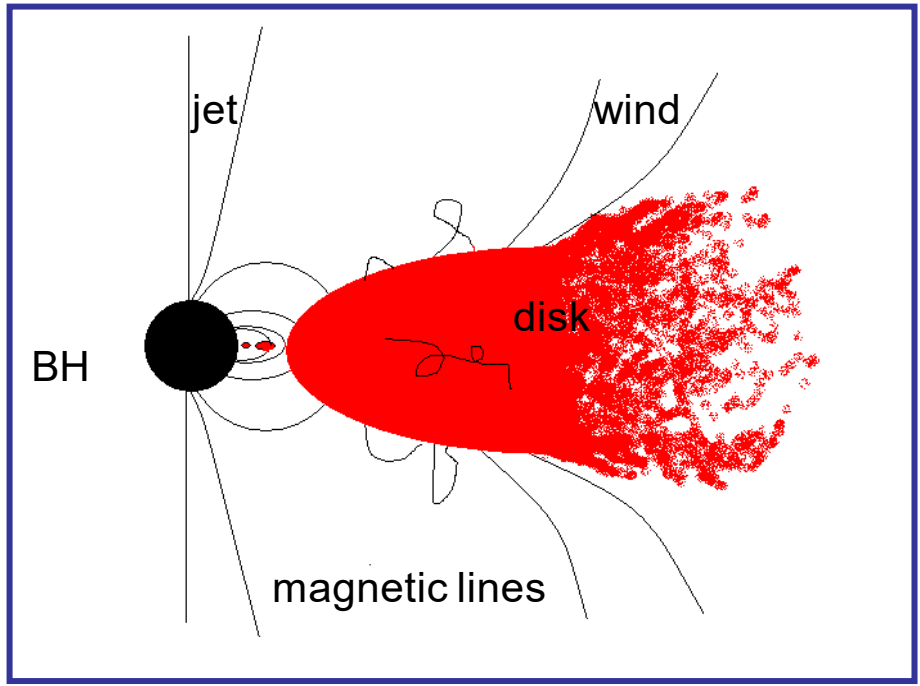

**Figure 13:** Wind of matter emitted by the accretion disk, for very bright quasars, near the Eddington limit. The wind is generated by the radiation pressure, aided by the magnetic pressure. The emission of the jet is closer to the black hole, and takes place, for example, per the Blandford-Znajek mechanism (see previous chapters).

The presence of significant radio jets (radio loud quasars) is anti-correlated with the brightness of the quasar. More precisely, the ratio between radio and optical brightness R = Lrad/LB is a decreasing function of the Eddington ratio λ = Lacc/LEdd (see Ho 2002, Figure 14). This is verified by both categories of AGNs, "radio-loud" and "radio-quiet", even if the ratio R is always higher for the first sequence. A saturation of the parameter R occurs at low accretion rates, for λ < $10^{-3}$, for low-luminosity AGNs, such as FRI among the radio sources, or even Seyfert galaxies and LINERs.

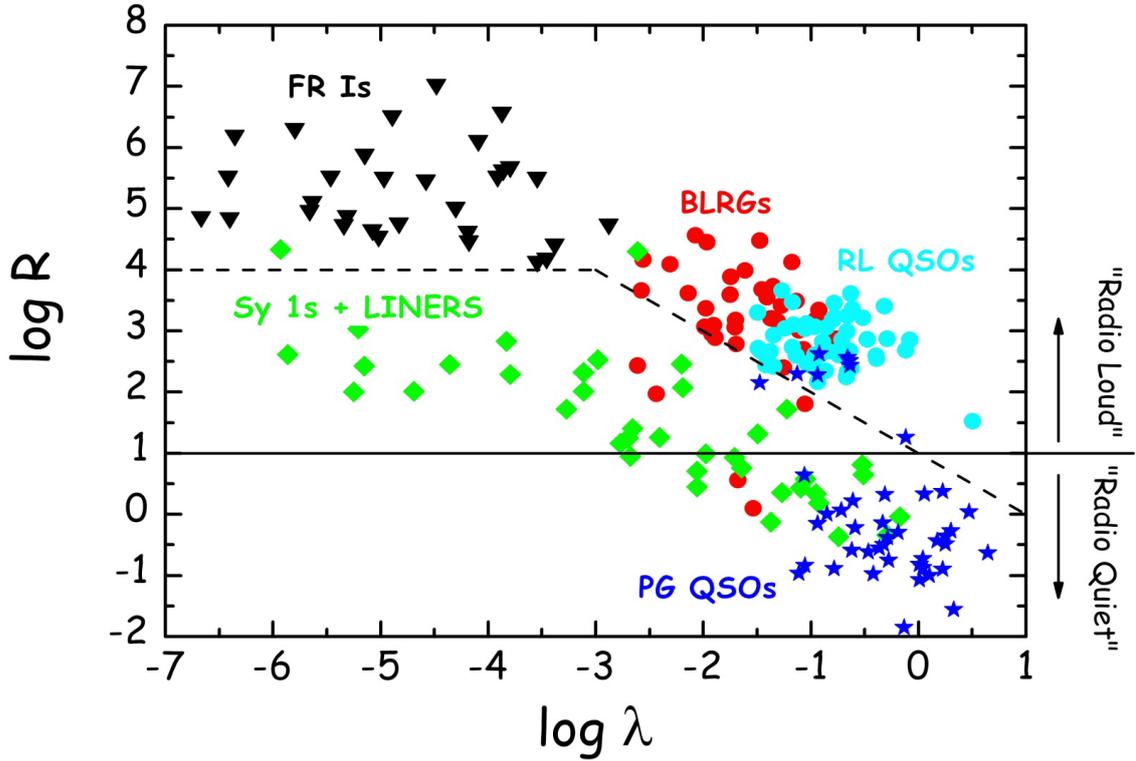

**Figure 14:** Anti-correlation between the ratio R of the radio and optical luminosities R = Lrad/LB, and the Eddington ratio of the accretion rate λ = Lacc/LEdd. The various colors correspond to the various AGN categories: blue the brilliant "PG, Palomar-Green" radio-silent quasars, the radio-loud or FRII quasars are in blue-cyan. Weaker radio sources (FRI) are in black, Seyfert and LINERs in green. In red are the BLR galaxies with broad lines. According to Sikora et al. (2007).

This anti-correlation can be explained by the various phases of black hole accretion, depending on its accretion rate compared to the Eddington rate (e.g., Fender et al 1999). In the very active, high-luminosity phase, matter forms a thin accretion disk, which radiates efficiently, and electrons are not emitted. The disk emits mostly soft X-rays. In the weak accretion phase, the disk is truncated until it no longer exists at all. Instead, the hot material, which does not radiate effectively, forms a plasma ring; the relativistic electrons then transform the soft X-rays into hard X-rays. This is why there is usually an anti-correlation between brightness and X-ray energy. In the low-light phase, radio jets can form, starting at the corona, which is the base of the jet. The corona favors the accumulation of magnetic flux near the black hole, thanks to its 3D geometry (Sikora & Begelman 2013). It would be favored by an episode of gas accretion or fusion between galaxies. It is thus explained that the hard X-rays are correlated with radio emission. Likewise, the most powerful radio sources are elliptical galaxies, formed by the fusion of spiral galaxies.

a. **The radiative or quasar mode**

For the radiative feedback mode, simple models have attempted to predict the ejection effectiveness of matter, based on the characteristic shock waves, well known in supernova models. Figure 15 schematically shows the different spheres, adiabatic (reverse) shock wave, contact discontinuity between wind and the interstellar medium, and shock wave advancing in the galaxy.

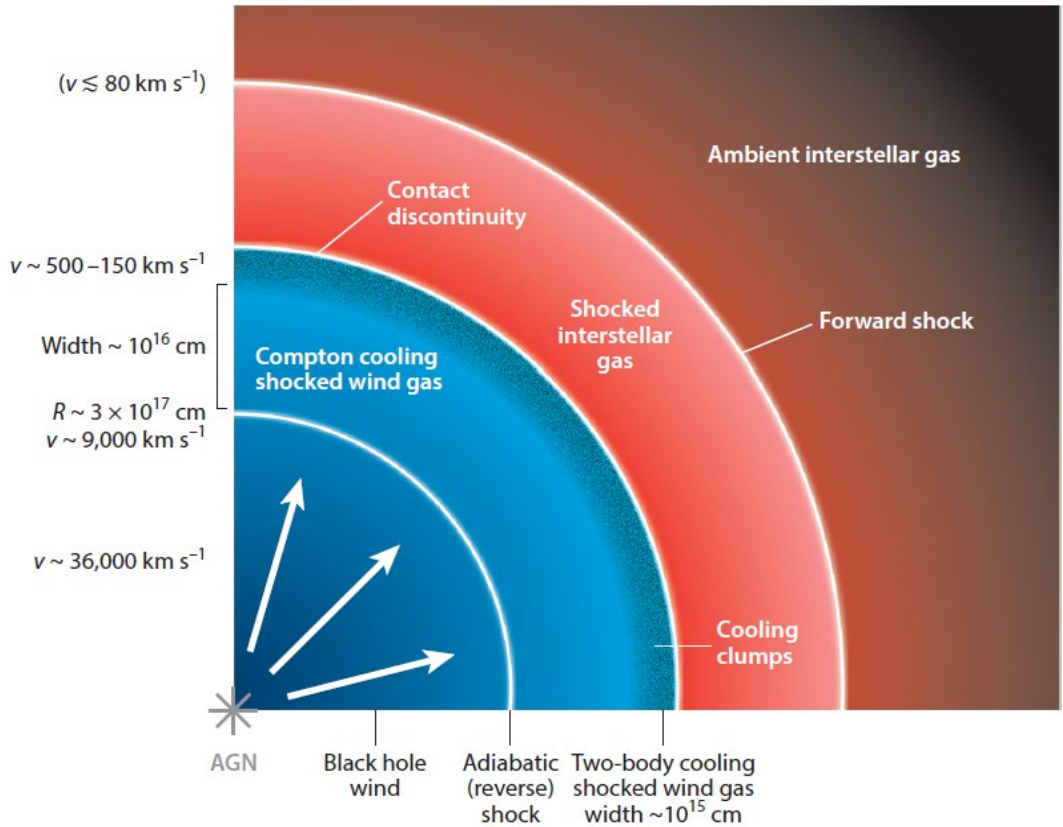

**Figure 15:** Diagram of the wind emitted by a high luminosity AGN, by radiation pressure, in the quasar radiative mode. A flow of plasma at very high speed (~c/10) produces a shock wave in the ambient interstellar gas. A reverse shock returns to the quasar, with the contact discontinuity surface between the two, which delimits the interstellar gas which has already undergone the shock wave, consisting of wind cooled at a more moderate speed.

Winds of almost relativistic speed near the black hole were observed by their iron lines in X-rays (Tombesi et al 2011). These streams of matter are called Ultra-Fast Outflows (UFOs). Their speeds are detected thanks to the lines of the highly ionized iron Fe XXV/XXVI in absorption. The speeds are typically c/30 to c/10.

Further in the galaxy, molecular gas flows are observed, at speeds of several hundred to a thousand km/s (e.g., Feruglio et al 2010, Cicone et al 2014). These observations were at first a surprise: how to explain that molecules remain in the waves? The shock waves should have destroyed the molecules. The gas is heated in the shocks to temperatures of $10^6$ to $10^7$K, enough to photo-dissociate hydrogen molecules and their tracers. However, the already shocked molecular gas, between the contact discontinuity and the shock wave, cools very efficiently, in a time scale well under a million years (Zubovas & King 2014). It becomes multi-phasic, with a phase fragmented into dense clouds, by Rayleigh-Taylor instabilities, which induces the formation of stars. This formation sometimes induces a luminosity comparable to that of an AGN, if it is greater than a rate of $100M_\odot$/year! In this case, it becomes very difficult to distinguish molecular flows due to starburst or AGN. The criterion of velocity can sometimes be used to decide. AGNs can eject gas at very high velocities, greater than the escape velocity, which is rarely the case with starburst galaxies (except for dwarf galaxies).

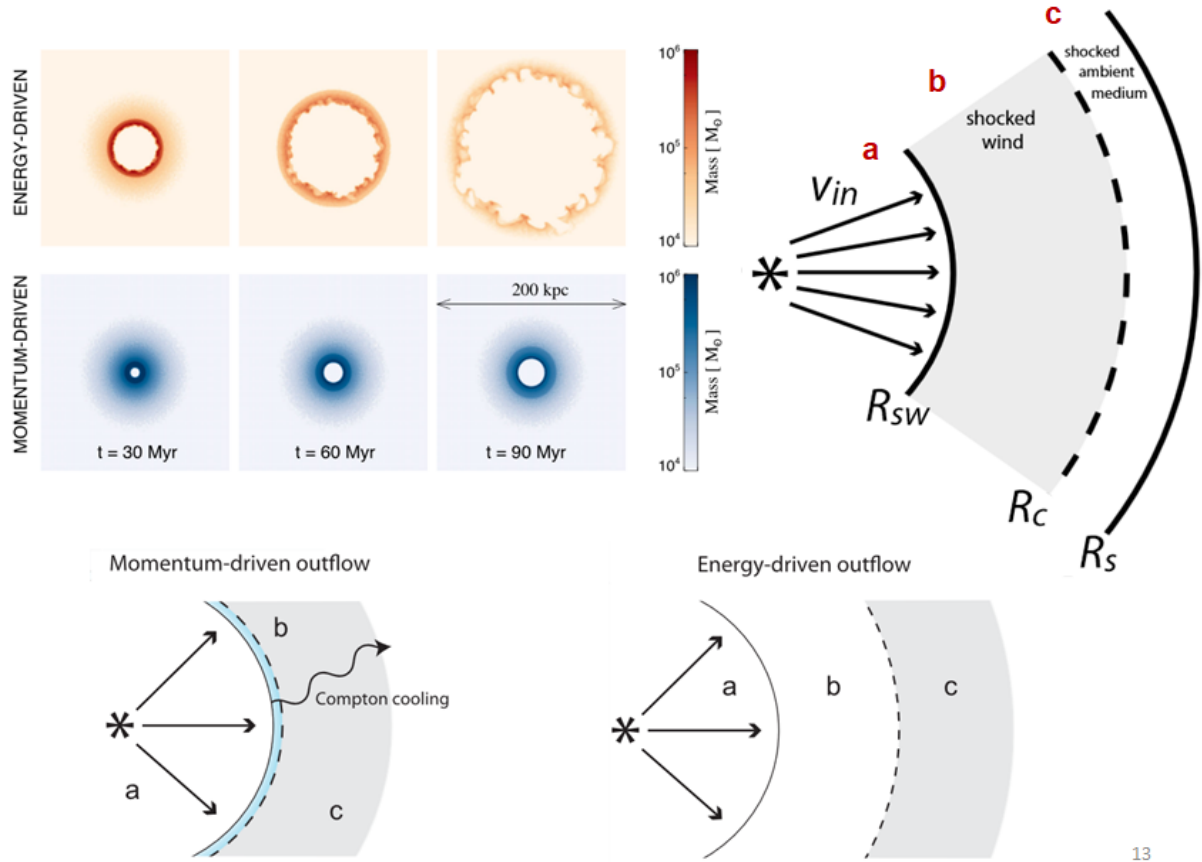

**Figure 16:** Diagram (top right) of the various outflow phases around an AGN in the radiative phase (quasar mode). $R_c$ is the discontinuity surface, $R_s$ the shock in the interstellar medium, $R_{SW}$ the reverse shock towards the relativistic wind (according to Faucher-Giguère & Quataert 2012). Zone "a" is characterized by the inflow velocity $V_{in}$, then zone "b" of the shocked wind, which can be considerably reduced in volume, if the cooling is effective, as in the hypothesis of a momentum conserved flow, which does not conserve energy. The "b" zone remains important, in the case where energy is conserved, as shown by the simulations of Costa et al 2014. At the top, in orange, the gas shell extends very far radially, in the case where energy is conserved. At the bottom, in blue, the extension of the gas is much smaller, because energy is not conserved. Each panel corresponds to a 200kpc square, for a black hole with mass $10^8 M_\odot$ which emits the Eddington luminosity. The stages of the simulation are 30, 60 and 90 million years.

There are several cases, as shown in Figure 16. Depending on whether the cooling in zone "b" of the shocked wind is effective or not, the shock energy is conserved or not, and how far the interstellar gas shell is ejected. If the energy is conserved, then we can observe a momentum in the molecular gas much greater than the initial momentum in the AGN wind. This case occurs for very fast winds, i.e., $V_{in}>10,000$km/s, with few radiative losses (Faucher-Giguère & Quataert 2012). The molecular gas velocity $V_s$, and flow $dM_s/dt$, then receives a momentum boost. If there is almost conservation of energy, to a factor of 2, we can write $dM_s/dt\ V_s^2 \sim 1/2\ dM_{in}/dt\ V_{in}^2$. And the quasar wind momentum $dM_{in}/dt\ V_{in}$ is approximately $L_{AGN}/c$. The molecular outflow momentum is $dP_s/dt = dM_s/dt\ V_s$, and that of

- 21 -

wind is dPin/dt = dMin/dt Vin, by conservation of energy we then have the ratio of moments: Ps/Pin = ½ Vin/Vs ; for Vin= 30 000km/s, and Vs~300km/s, the ratio Vin/(2Vs) can reach up to 50!

This explains why the molecular flow momentum is much greater than LAGN/c. Among the various molecular flows observed in recent years, statistics have shown that more energy is conserved. In fact, the flow momentum is on average 20 times greater than that of the AGN wind (see Figure 17, Cicone et al 2014).

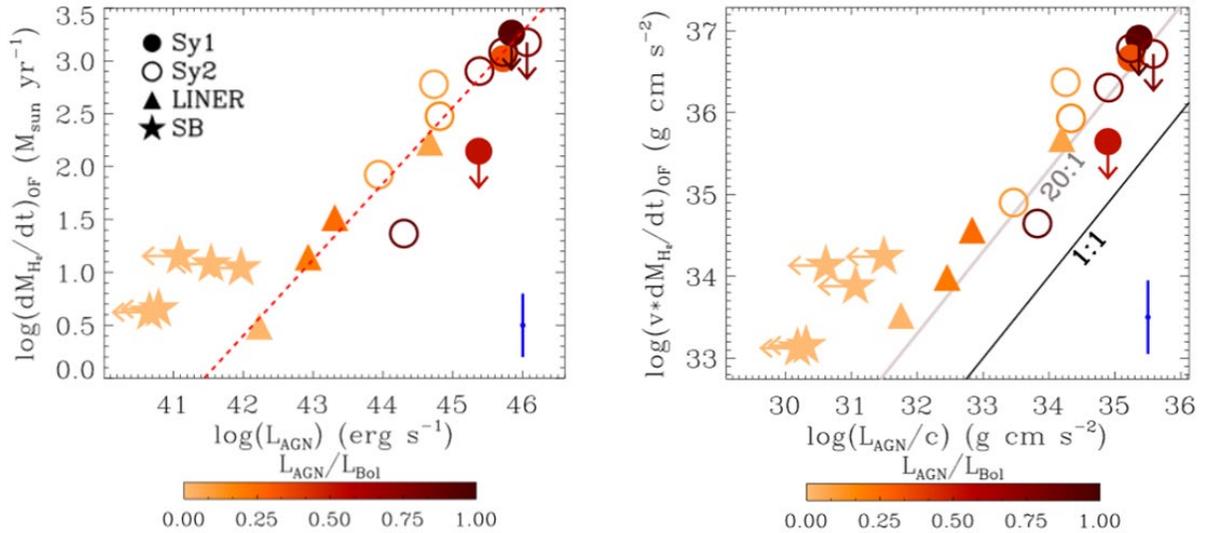

**Figure 17:** Left: molecular outflow rate, as a function of the AGN bolometric luminosity ($L_{AGN}$), for several objects, identified by the symbols: circles = Seyfert galaxies, triangles = LINERs, stars = starburst galaxies. The color indicates the relationship between the luminosity of the AGN and the bolometric luminosity of the host galaxy ($L_{AGN}/L_{bol}$), according to the color palette shown below. Only galaxies with an AGN verify the correlation (not starbursts). On the right, with the same symbols and colors, the molecular flow momentum is plotted, as a function of the quasar wind momentum $L_{AGN}/c$. According to Cicone et al (2014).

### b. The mechanical mode, or radio mode

When AGN is in a low-luminosity phase, the only energetic manifestation is that of the radio jet, if it exists. The jet can drag out interstellar gas, if the coupling with the disk is sufficient. This is often the case, because the jet is not aligned perpendicular to the plane of the galaxy, as one might think. Instead, the jet is perpendicular to the accretion disk, but the latter is quite independent and decoupled from the orientation of the large-scale disk. The sizes are very different, the molecular tori are of the order of parsec, the galactic disks of the kilo parsec, and the timescales the same (dynamic times different by a factor of 1000). Even gravitationally, the potential felt by the accretion disk and the molecular torus is quasi-spherical, Keplerian under the influence of the central black hole, and does not know the orientation of the large-scale disk. Very often the jet is observed sweeping the gas of the galaxy (e.g., Garcia-Burillo et al 2014, Dasyra et al 2015).

The radio mode is characterized compared to the quasar mode, not only by the luminosity of the AGN, but also by the collimation of the molecular outflow. Sometimes the latter looks like an extremely fine jet (Aalto et al 2016), and only a radio jet can cause it, even if it has not yet been observed. On the other hand, the impact of a highly collimated radio jet does not seem likely to be very significant, if its action is only to dig a channel in the interstellar medium and thus escape the galaxy. However, it is not, because step by step, the jet heats all the surrounding gas, and by Kelvin-Helmholtz instabilities, has a cross section much larger than its own. It then forms a cocoon, the impact of which is considerable on the galaxy, both in energy and amount of momentum transferred. The simulation of Figure 18 shows the formation of this cocoon, in a fractal medium representative of the interstellar medium, and its porosity (Wagner & Bicknell 2011).

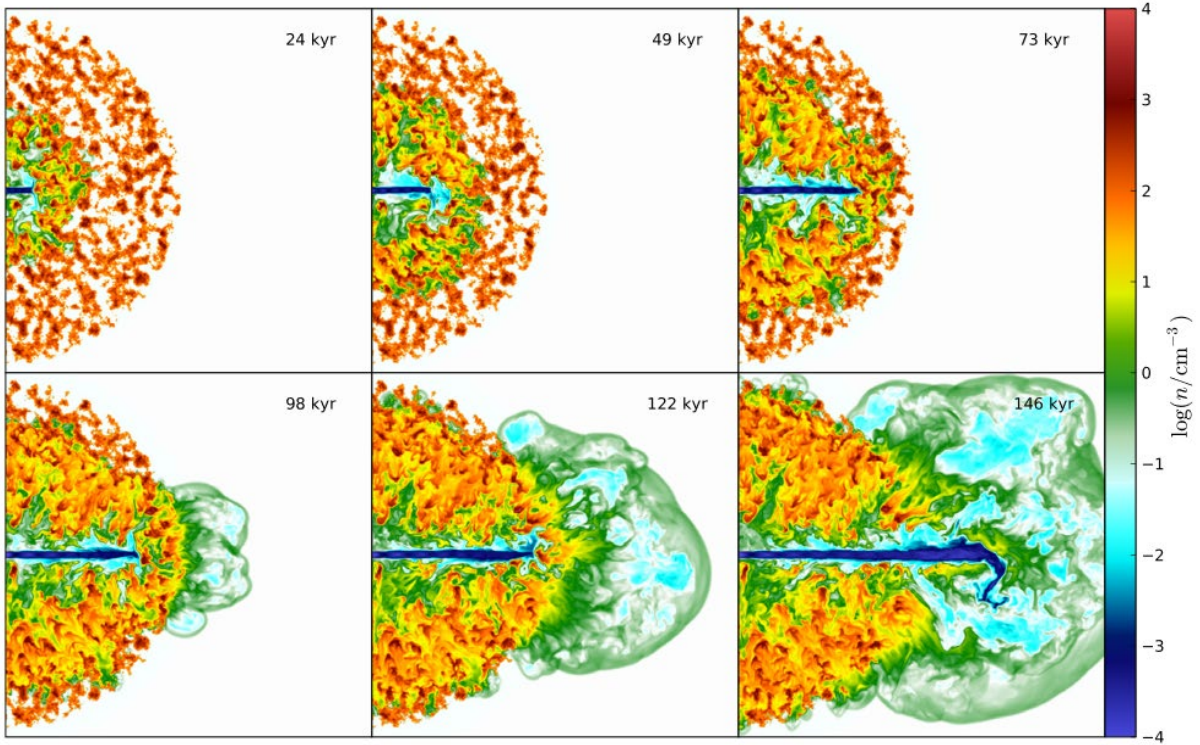

**Figure 18:** Simulation of the development of a plasma jet in a fractal interstellar medium. Each panel is 1kpc on a side. The color indicates the density of the medium, according to the palette to the right. The fractal structure consists of fragments between 2pc and 1kpc in size. According to Wagner & Bicknell (2011).

## 3  Black hole fueling processes

### 3.1  Problem of angular momentum, stellar processes

To feed the central black hole, the matter must lose a lot of angular momentum. As the latter is proportional, per unit of mass, to the velocity and to the radius, and the velocity varies very little, it is the radius that determines the orders of magnitude. For matter near the center (~ 1kpc), it will be necessary to arrive on the accretion disk, whose radius is typically $10^{-2}$ pc, therefore a loss of a factor $10^5$. What is the inflow of matter necessary for a luminous quasar? A quasar can emit up up to L=$10^{13}$ L$_\odot$, i.e., $10^{39}$ W ($10^{46}$ erg/s). With a radiation efficiency of $\varepsilon$=10%, the luminosity is written L = dM/dt $c^2$ $\varepsilon$, you therefore need an accretion rate of dM/dt = 1.7 (0.1/$\varepsilon$) (L/$10^{39}$ W) M$_\odot$/year. Assuming typical duty cycles of 100 million years, the black hole would have to swallow $2\ 10^8\ M_\odot$, meanwhile, which is a notable fraction of the gas in a galaxy. This matter is not in the immediate vicinity, and it will take a dynamic mechanism to exchange the angular momentum at the kpc scale in the galaxy.

To bring in the gas in such a short time, and cause it to lose its angular momentum, torques are required, and therefore tangential forces, produced by disk disturbances. If the disk remains axisymmetric, the gravitational forces are only radial, and there are no torques. Asymmetries can be bars, spirals, interactions between galaxies. The formation of these disturbances has been described in the "Galaxies" volume of this series; in this chapter we will describe and quantify the angular momentum exchanges.

First, let us consider the matter in the immediate environment of the black hole, which could initially feed it: the stars.

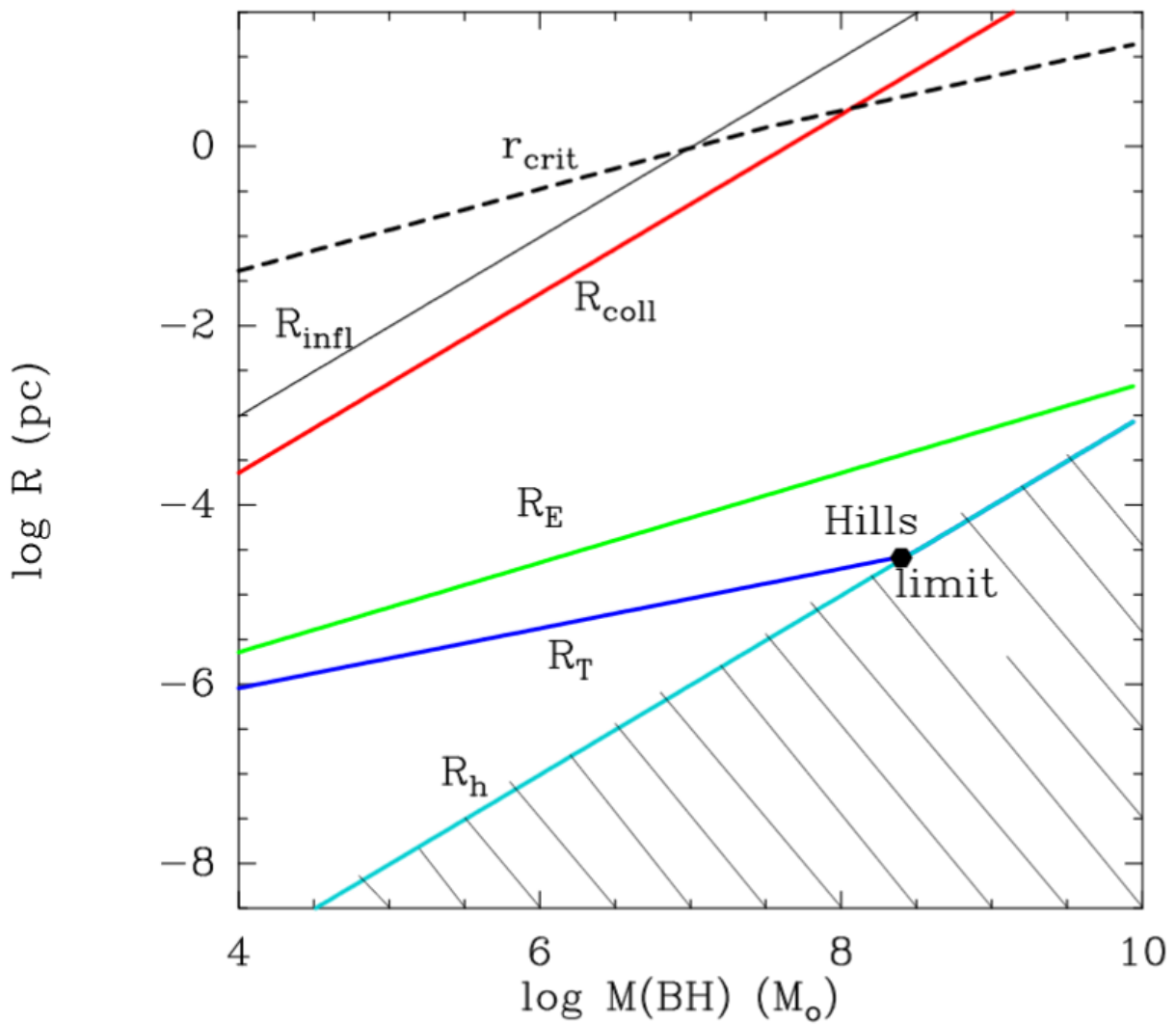

**Figure 19:** Characteristic radii corresponding to different physical phenomena, depending on the mass of the black hole. From top to bottom, $R_{infl}$, the radius of the black hole sphere of influence, $R_{coll}$, the collision radius, $R_E$, the Eddington radius, $R_T$, the tidal radius, $R_h$, the black hole horizon or Schwarzschild radius (see definitions in text). The loss cone effects for stars are significant within the critical radius $r_{crit}$ (dotted lines).

Stars lose mass over their lifetime, on average for a star cluster, at a rate of $10^{-11}$ $M_\odot$/year per $M_\odot$ of star. If there is a nuclear star cluster in the few central parsecs, then it is possible to have enough gas that has already lost its angular momentum to form the star cluster. The formation of this nuclear cluster could be due to a merger of galaxies, capable of forming a cluster of $4\,10^9$ $M_\odot$ and therefore provide $\sim 10^7$ $M_\odot$ in one activity cycle almost simultaneously with the starburst (Norman & Scoville 1988). However, in this starburst formation, the direct gas supply may be dominant. The stars could also help fuel the black hole when there is no gas. They can be destroyed by tidal forces, reduced to gas, and then swallowed by the black hole accretion disk. For this, the mass of the black hole $M_{BH}$ must be less than the Hills mass, of $3\,10^8$ $M_\odot$. This limit arises from the fact that the average density $<\rho>$ of a black hole inside its horizon (of radius $R_h = 2\,G\,M_{BH}/c^2$) is a decreasing function of its mass: $<\rho> \propto M_{BH}/R^3 \sim 1/M_{BH}^2$. This density becomes equal to the average density inside a star of the main sequence, for $M_{BH} = 3\,10^8$ $M_\odot$. For more massive black holes, stars are no longer destroyed by the tidal effect outside the horizon, but inside, and no longer contribute to AGN activity (Hills 1975). It should also be taken into account that stars in the vicinity of a black hole can be "swollen" or "bloated", therefore more fragile and sensitive to mass loss. The stars could as well be giants. There is also another process, which could fuel the black hole: collisions between stars, in the dense stellar cluster existing in the nucleus. The velocities of stars as they approach the black hole are relativistic, and collisions can tear off a lot of mass, or even destroy stars (Rauch 1999). Collisions also have the advantage of redistributing angular momentum, and thus refilling the loss cone, through the distribution of low angular momentum stars, which have already been swallowed by the black hole. Collisions between stars are now a preferred way to form intermediate-mass black holes in dense star clusters (Reinoso et al 2018, Sakurai et al 2019).

The processes to fuel the super-massive black hole with stars alone, are necessary for AGNs active at z = 0 in galaxies with no gas available. It is useful to quantify them to know some characteristic scales, illustrated in Figure 19. The radius $R_{infl}$ corresponds to the black hole sphere of influence, $R_{infl} = GM_{BH}/<V^2>$, already defined at the beginning of this chapter, depending on the velocity dispersion of the bulge $<V>$. The collision radius $R_{coll}$, is the one below which collisions release gas, i.e., the velocity of stars around the black hole $(GM_{BH}/r)^{1/2}$ is comparable to the escape speed $v_*$ of a typical star with mass $M^*$ and radius $r^*$, $v^* = (GM_*/r_*)^{1/2}$. The Eddington radius $R_E$ is the radius below which a star receives more energy than its own Eddington luminosity, $R_E \sim r_*(M_{BH}/M_*)^{1/2}$. From this moment, their envelopes can evaporate, or else form "bloated" ("swollen") stars, more fragile to mass loss. The tidal radius $R_T$ is that below which a star can be destroyed by tidal forces from the black hole, it can be calculated as the Roche radius, where the density of a star equals the average density of the black hole: $R_T \sim r_*(M_{BH}/M_*)^{1/3}$. La Figure 19 shows quite well the Hills massl when $R_T$ becomes equal to the horizon of the black hole. There is then no more activity in the nucleus, when the black hole swallows a star, on the other hand its mass still grows!

## 3.2    Fueling from gas, gravitational torques

To evacuate the angular momentum of the gas outwards, torques are required. Could these be viscous torques? Interstellar gas in a galactic disk actually has very low viscosity. It is very diffuse (atomic gas of average density 1 particle per $cm^3$), or else fragmented into dense molecular clouds, which collide barely once or twice per rotation. Clouds cannot be viewed as a fluid, but as ballistic particles, with some dissipation due to collisions. In the thin accretion disk around the black hole, there are very significant viscous torques. Viscosity is modeled as a force proportional to pressure. The magnitude of the viscous torques is such that they can reduce gas to a radius of 4pc in a billion year. Beyond that, the timescales will be much too large for viscosity to be efficient (e.g., Pringle 1981). Viscous torques are negligible for galactic disks, at the kilo parsec scale, but the disks are subject to gravitational instabilities. The instabilities will produce disturbances, such as spiral arms, bars. The instabilities heat the disk, create the dispersion of streaming motion velocities, which stabilize the stars (Toomre parameter Q >> 1). The gas is dissipative and radiates this energy. The Toomre parameter reverts to 1 for gas, which may be subject to other instabilities, and so continuously. These instabilities, constantly renewed, transfer mass and angular momentum, and are at the origin of an effective kinematic viscosity (Lin & Pringle 1987).

The nature of spiral and barred instabilities has been described in the Galaxies volume (chapter 4). The stellar bar of a galaxy is a density wave that can generate spiral waves in the gas component. The stars follow in their orbits families of periodic orbits that are parallel or perpendicular to the bar, changing 90° with each Lindblad resonance. The gas flow lines also gradually rotate 90° at each resonance, following a spiral wave. The gravity torques exerted by the bar on the gas spiral can be summarized in Figure 20. They change sign at each resonance.

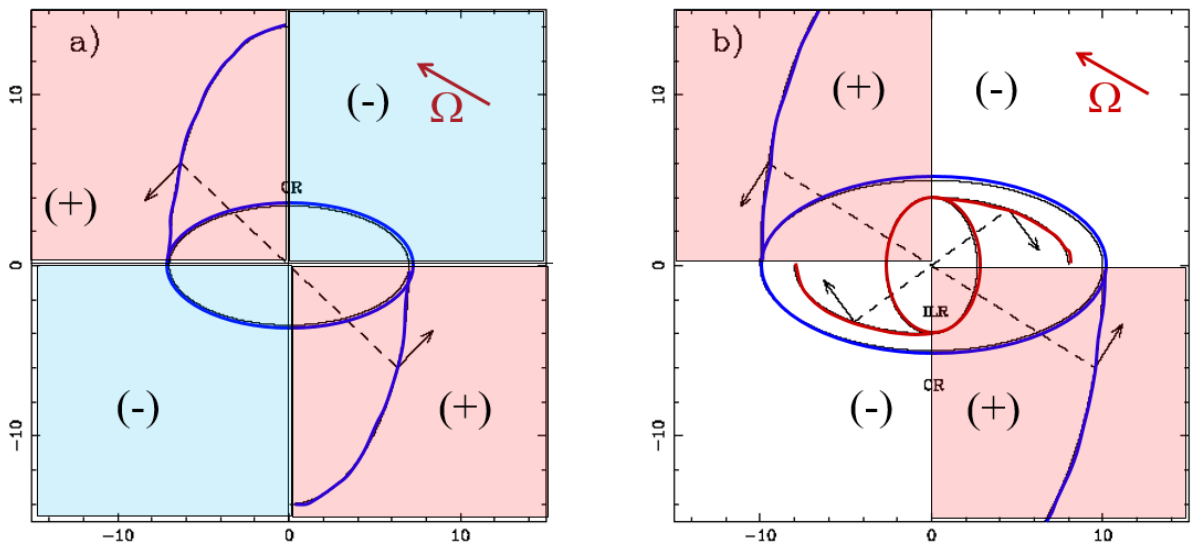

**Figure 20:** The stellar bar (here horizontal) divides the plane of the disk into four quadrants: in pink, the gravitational torques of the bar on the gas are positive, and in blue, the torques are negative. The direction of the torques is oriented by the direction of rotation in the galaxy (here the direct direction). The spiral arms generated by the barred potential on the gas are in the pink quadrants outside the corotation (ends of the bar), and in the blue quadrants between corotation and Internal Lindblad Resonance (ILR).

It is evident from the nature of the gravitational torques that the gas will be driven from the corotation outwards, where it will accumulate at the Outer Lindblad Resonance (OLR). At resonances, the gas settles in rings, circular or elongated, which are symmetrical about the bar: thus, no torque is applied to it, and the gas accumulates to form stars. Inside the corotation, the torque is negative, and the gas will accumulate in an internal resonance ring or ILR.

To quantify the time scales of these angular momentum transfers, it is possible to calculate the gravitational torques directly on the observed galaxies. The gravitational potential of a spiral galaxy, mainly due to stars, is obtained by projecting a near infrared image. The forces are calculated at each pixel of the disk by derivation, assuming the disk is thin, isothermal, of a constant height scale equal to $\sim 1/12^{th}$ of the radial scale of the image (Garcia-Burillo et al 2005). On the torque map, which is broken down into four quadrants around the bar, the contours of the molecular gas overlap, as shown in Figure 21. It is easy to see if the gas is concentrated in the positive or negative regions. By taking the azimuthal average, we obtain at each radis the angular momentum transfer rate.

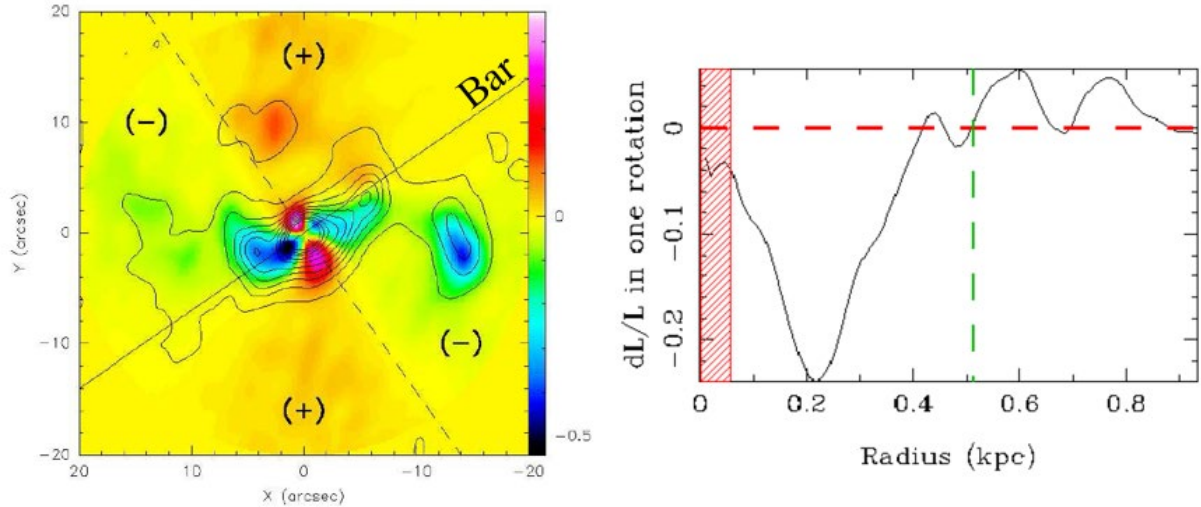

**Figure 21:** Calculation of the gravitational torques in an observed spiral galaxy. Left: color map of torques in the de-projected galaxy. The major axis of the galaxy is horizontal. The bar separates the space into four quadrants of positive (red) or negative (blue) torques. The contours represent the observed molecular gas density. On the right: from the 2-dimensional map of the torques (figure on the left), the azimuthal average of the torques exerted on the gas, weighted by the density of the gas in each pixel, is calculated for each radius. The torque is equal to the change in angular momentum dL/dt, and the unit of the vertical axis is chosen as the relative change in L in one rotation, at each radius. The torque is negative at the center, and will transfer 30% of the angular momentum of the gas in one rotation. According to Casasola et al (2011).

The calculation carried out on around thirty nearby galaxies, with an active nucleus, but of low luminosity (Seyfert or LINER) showed that at the scale of ~100pc, only a third of the galaxies were in the active nucleus feeding phase (Garcia -Burillo & Combes 2012). The rest of the time, the gas is stalled at the ILR and forms stars in a ring. From the fraction of negative torques, we can deduce a fairly short feeding phase of black holes of around $10^7$ years. Once the mass has accumulated sufficiently towards the center and into the ring, the next step is the decoupling of a secondary bar, or a nested nuclear structure. Mass accumulation increases the rotational velocity at the center and the rate of precession $\Omega-\kappa/2$, which is close to bar pattern speed. Two ILR resonances occur, weakening the primary bar, as the periodic orbits between the two ILRs are perpendicular to the bar. A faster secondary bar decouples in the center, inside the smaller ILR, as shown by numerical simulations. A certain number of nested structures can even follow one another, of symmetry m = 2 and m = 1 closer to the black hole (Keplerian potential, see Hopkins & Quataert 2010).

## 3.3 Sphere of influence and molecular torus

The series of nested structures must certainly come to an end at the scale of the molecular torus, of size 1-10pc, the existence of which was suggested very early in the unification paradigm of AGNs (see chapter 1). It is only recently, with the very high resolution and sensitivity of millimeter interferometers (ALMA, NOEMA) that it has been possible to detect molecular tori and structures within the black hole sphere of influence.

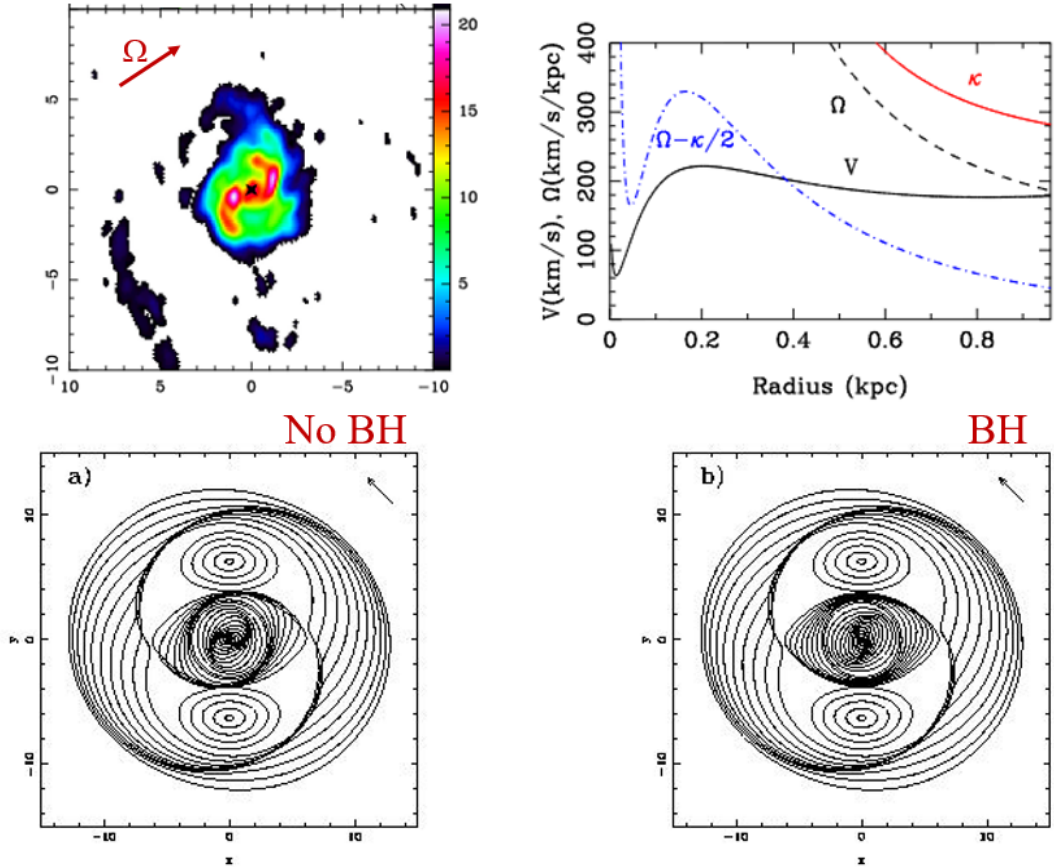

**Figure 22:** Detection of a "trailing" nuclear spiral at the core of NGC 1566 (Combes et al 2014): top left, map of the CO line (3-2) observed with ALMA, in the projected galaxy plane (axes are in arcseconds, 1" = 35pc). Top right, rotational velocity V, $\Omega=V/r$ and epicyclic frequency $\kappa$ in NGC1566. The blue curve $\Omega-\kappa/2$ is the precession rate of elliptical orbits. This rate goes up in the black hole sphere of influence to within 50pc. Bottom: representation of the gas flow lines in the bar rotating frame (the bar is horizontal). At the corotation, the gas travels only through an epicycle at the Lagrange points L4 and L5, on the vertical of the center. In the absence of a black hole, the orbital precession velocity $\Omega-\kappa/2$ does nor increase towards the center, and the nuclear spiral structure is "leading". It is "trailing" with a black hole. According to Buta & Combes (1996).

As one enters the black hole sphere of influence, the precession velocity of the elliptical orbits, which drives the gas streamlines, increases, as the frequencies $\Omega$ and $\kappa$ vary in a Keplerian manner as $r^{-3/2}$. When the gas is precipitated towards the center by the gravitational torques, its rate of precession first decreases, which causes the spiral to proceed in the leading direction, then it increases and the spiral reverses in the trailing direction. This has been observed for the galaxy NGC 1566, as shown in Figure 22 (Combes et al 2014). The winding sense of the spiral controls the sign of the gravitational torques, as shown in Figure 20. Without the black hole, the spiral is leading and the torque is positive: the gas accumulates towards the ILR (inner resonance). On the other hand, in the presence of a black hole, the torque is negative, and the gas will feed the nucleus.

A similar trailing nuclear structure was also observed in NGC 613 and NGC 1808 (Audibert et al 2019). The precise calculation of the gravitational torques has confirmed the feeding of the black hole, with a timescale of around 4 million years (see Figure 23).

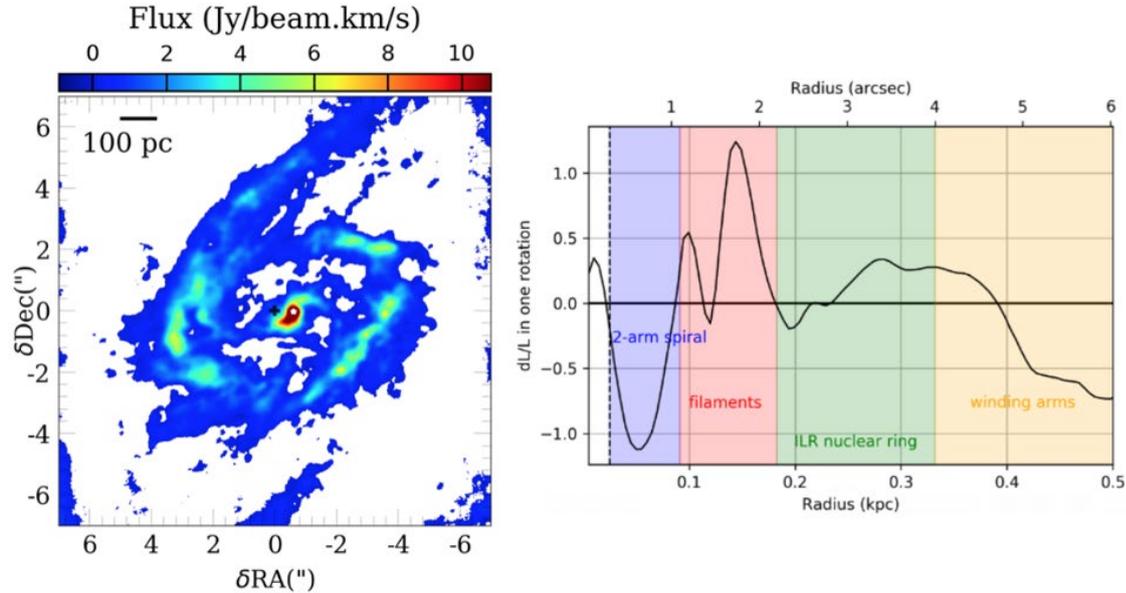

**Figure 23:** ALMA map of the CO line (3-2) at the center of NGC 613, showing the detection of a trailing nuclear spiral inside the ring at ILR resonance (1"=83pc). On the right, the calculation of the gravitational torques shows that inside 100pc, the gas loses its angular momentum in one rotation. According to Audibert et al (2019).

The nuclear spiral structures observed are the last before the molecular torus. Just inside these spirals, compact circumnuclear disks have been observed, which appear to be decoupled by their morphology and kinematics. Most often, their orientation and inclination on the sky plane are different from those of the disk of the host galaxy (Combes et al 2019). These components correspond to what is expected, to account for the obscuration of certain active nuclei, depending on their orientation on the line of sight. It is not clear from spatial resolution whether they are really torus-shaped with a molecular gas deficiency in the center, except in a few rare cases. The radius of molecular tori varies from 6 to 27 pc, and their mass from 0.7 to 3.9 $10^7$ M$_\odot$. The torus is generally centered on the continuum source detected in millimeter waves, which is essentially synchrotron emission from AGNs.

It is no surprise that the circumnuclear disk (or torus) is decoupled from the rest of the galaxy. Indeed, inside the black hole sphere of influence, when the potential becomes Keplerian and spherical, the gas no longer has any information about the orientation and flattening of the rest of the galaxy. However, there is still a memory of the original angular momentum of the gas. The phenomenon of star formation, though, and associated feedback (ejection of supernovae, fountain effect above the plane, and subsequent gas accretion) randomizes the new orientations of the gas (e.g., Emsellem et al 2015).

The molecular gas being observed in the black hole sphere of influence, in relative dynamic equilibrium, it is possible to deduce the mass of the black hole, by the kinematic method. This method is all the more welcome since the other methods do not give very precise results for low spheroid masses, and low black hole masses. Scattered measurements may reflect intrinsic dispersion, but more galaxies observed in molecular gas at high angular resolution are needed to conclude (see Figure 24).

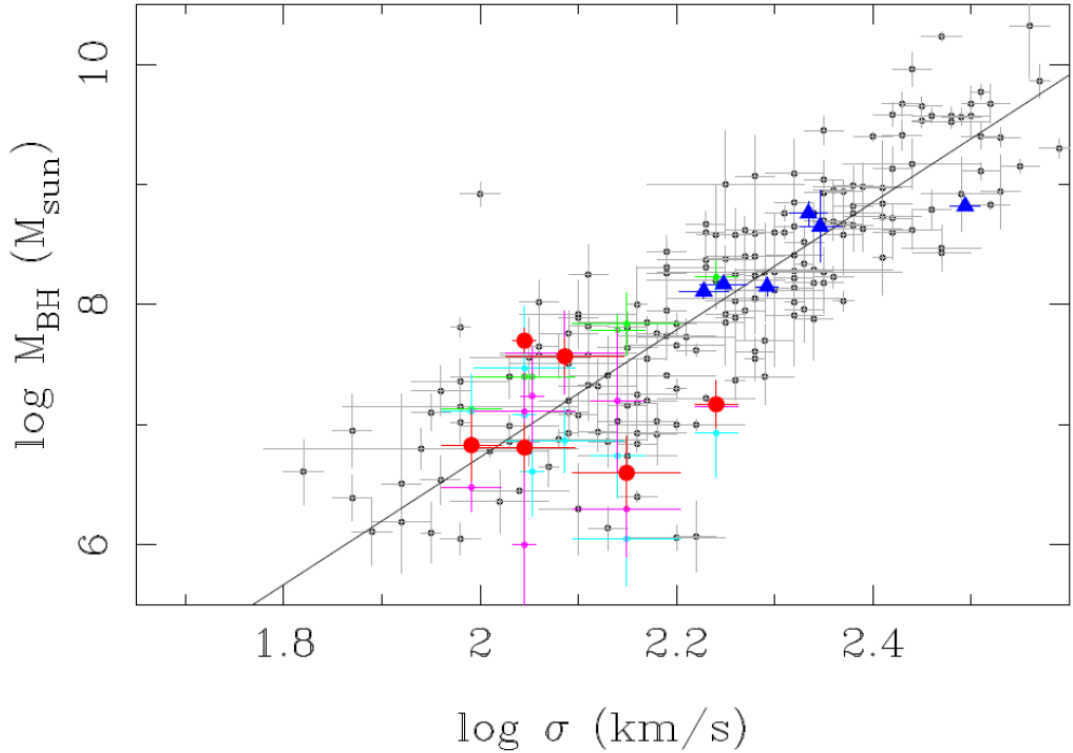

**Figure 24:** Measurement of the mass of black holes, by the molecular gas kinematics method in the black hole sphere of influence. Spiral galaxies, with low-luminosity AGNs, are red circles (Combes et al 2019), and the similar measurement in elliptical galaxies with higher spheroid masses is in blue triangles (WISDOM project, Davis et al 2018). The gray symbols come from the compilation of van den Bosch (2016), adjusted by a straight line of slope 5.35. Estimates of black hole mass in spiral galaxies using different methods (red circles) are also shown in green, turquoise and pink.

## 4 AGN feedback– Efficiency

### 4.1 The need for feedback

The feedback phenomenon is invoked in the standard cosmological scenario ΛCDM to avoid forming galaxies that are too massive. As shown in Figure 25, the mass function of the observed galaxies is very far from what might be expected, from the mass function of the dark matter halos, assuming that the baryon fraction is equal to the universal fraction ($f_b$=17%) in each halo. The observed baryon deficit is particularly significant for small masses, where the feedback from supernovae is supposed to be very efficient, given the shallow depth of the potentials of these dwarf galaxies. For the most massive galaxies, feedback from supernovae is not efficient, and the role of AGNs is critical. The simulations succeed in suppressing the formation of very massive galaxies and it remains to verify the efficiency of this feedback in the observations.

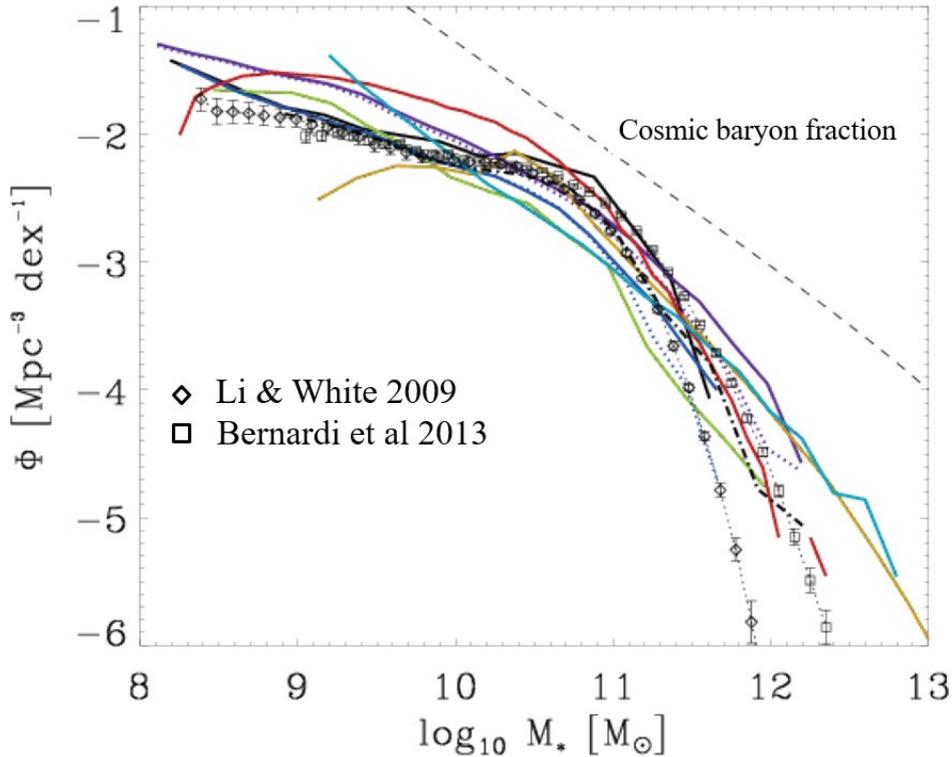

**Figure 25:** Observed mass function of galaxies, rhombuses and squares with error bars (Li & White 2009, Bernardi et al 2013) compared with cosmological simulations using the feedback effects of star formation and AGNs, calibrated on the observations (various color curves, corresponding to several simulations). The top dashed curve is the theoretical mass function, based on the mass function of dark matter halos, assuming the universal baryon fraction, without feedback. According to Naab & Ostriker (2017).

Feedback from AGNs is particularly necessary for the most massive galaxies. However these are the ones that have the most massive black holes too, and for which it is not certain that outward manifestations of energy are possible. Indeed, a quasar radiates efficiently from a thin accretion disk, which can no longer exist beyond a certain black hole mass, which we will define now. Indeed, the outer limit of the accretion disk, at the origin of the AGN emission, is a function almost independent of the mass $R_{disk} \sim 0.01$ pc $\sim 2000$ AU. Beyond that, the disk cannot subsist, the medium becomes self-gravitating and forms stars. On the other hand, the radius of the horizon is proportional to the mass of the black hole $M_{BH}$, $R_h \sim 20$ ($M_{BH}/10^9$ $M_\odot$) AU, and the last stable orbit is 3 $R_h$ for a non-rotating black hole and 0.5 $R_h$ for maximum rotation (a = 1). In any case the disk can only exist between this last stable orbit, and $R_{disk}$, and no longer exists for a black hole mass greater than $5 \cdot 10^{10}$ $M_\odot$, assuming typical parameters (King 2016). Black holes can still increase their mass, but they will be invisible, with no feedback phenomenon possible.

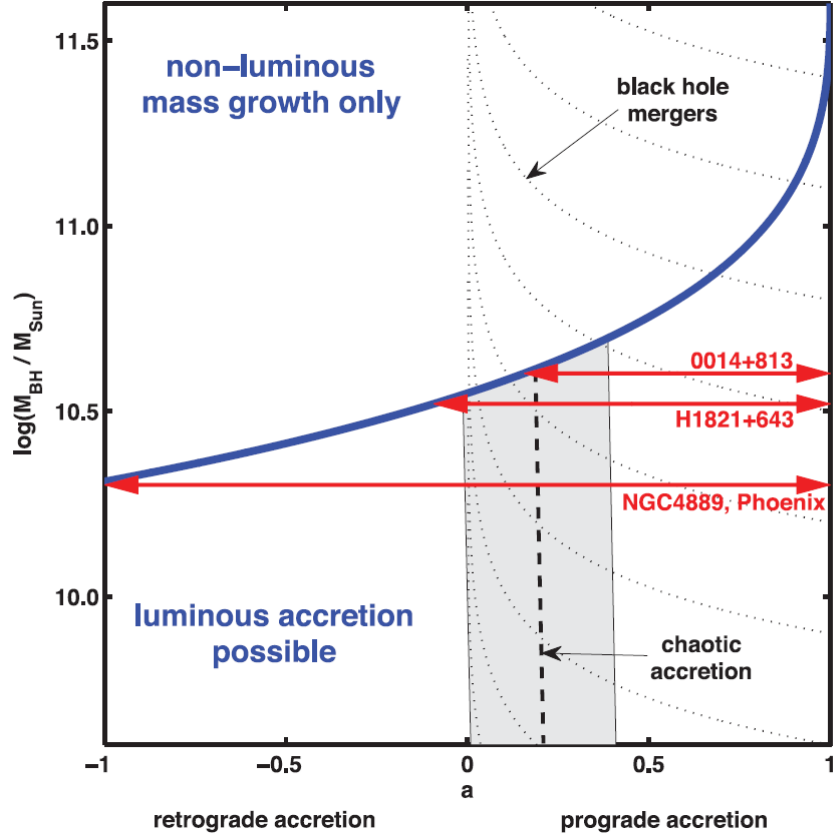

**Figure 26:** Calculation of the maximum mass that the black hole of an AGN can have (curve in blue). A black hole can always increase its mass beyond, but this will be done without luminosity (therefore without feedback). The abscissa is the dimensionless parameter 'a', characterizing the spin of the black hole. The parameter a is between 0 and 1 (maximum rotation). Here negative 'a' values indicate retrograde accretion of matter with respect to the black hole spin. Three of the highest measured black hole mass values are shown in red. We see that the two highest values can already constrain the spin of the black hole. The dotted lines show the statistical effects of black hole mergers on their mass growth. According to King (2016)

## 4.2 Prototype: cooling flows

Clusters and groups of galaxies are bathed in a very hot gas, visible in X-rays, the temperature of which can reach 100 million degrees for the richest clusters. This gas is in virial equilibrium in the cluster potential well, and the average timescale of its radiative cooling is greater than the Hubble time. Yet the cooling rate is proportional to the square of the density, because the radiation of the ionized gas occurs by collision between particles. At the center where the density is higher, the cooling time $t_{cool}$ becomes less than the Hubble time, typically $t_{cool} \sim 300$ million years. The gas must therefore cool, lose its pressure support and fall towards the central galaxy. The latter called Brightest Cluster Galaxy (BCG) is a very massive galaxy, often a cannibal, which swallows the satellites spiraling towards the center by dynamical friction. Most of the time it is also an AGN. Gas accretion rates up to 1000 $M_\odot$/year are predicted, but are not observed. The stars that this cold gas should have formed are not detected either. Many hypotheses have been proposed, such as the conduction of heat from the external parts towards the center to avoid its cooling, a blocking magnetic field, abundance anomalies, etc. However, since 2002, astronomers began to detect cold and molecular gas in the center of the clusters, corresponding to 10% of the expected cooling rate, and understood that this flow of gas was at the origin of the activity of the BCG nucleus. The resulting AGN is able to limit, and moderate the cooling rate, by the associated feedback phenomena (Edge 2001, Salomé & Combes 2003, Peterson & Fabian 2006).

The impact of AGN on hot gas cooling is via the radio mode. In relaxed, cool-core clusters, 70% of BCGs are radio sources, while there are only 20% in other clusters. The latter do not cool down, because they are dynamically disturbed by collisions between groups and accretions of sub-clusters. It is possible that all the clusters spend some time in the relaxed and disturbed phases, cool-core or not. In the case of feedback, the radio jet of the central elliptical galaxy digs quite remarkable cavities in the hot cluster gas, up to 100kpc from the center (see Figure 27). The cavities fill with plasma from the jet, forming a cocoon. This relativistic and non-thermal gas (synchrotron emission) is less dense than the hot gas in the cluster (thermal emission). They are in pressure equilibrium, but the less dense rises by Archimede forces. Cavities are bubbles that move, rise through buoyancy, and move away from the center. They occur in each phase of AGN activity, the jet of which is intermittent. This is why it is possible to see several cavities or bubbles in the hot gas (up to three sets, because then the bubbles disperse). The periods of activity of the jet are around 40 million years, similar to the rise time of the bubbles up to 20kpc.

The cold molecular gas (T ~ 20K) is observed far from the center, concentrated just outside the wall of the cavities (see Figure 27). It is possible that hot low entropy gas has been entrained by the bubbles and cools just above the bubble wall, where the gas is compressed, and the cooling time $t_{cool}$ becomes about 10-20 times the free-fall time $t_{ff}$ (McCourt et al 2012). Alternatively, it can be considered that the gas infall time $t_{infall}$ is much greater than the free fall time, due to the turbulence and the hot gas entrained by the bubbles, and that $t_{cool}/t_{infall}$ ~1 is the factor triggering the cooling (McNamara et al 2016). Simulations have shown that when $t_{cool}/t_{ff}$ drops below 10-20, the hot gas is subject to thermal instabilities, which make the gas turbulent and inhomogeneous. The density can be higher than average in places and cause cooling (Gaspari et al 2012). As soon as the gas cools, it fragments, becomes dense and molecular, loses pressure support and should fall in free fall, in the form of filaments. In fact, the gas velocities observed by the CO lines ($H_2$ tracer) show a much lower speed, showing that the gas is slowed by the dynamical ram pressure of the hot gas and/or the magnetic field (Olivares et al 2019). The filaments are spectacular in ionized gas, mapped by Hα (Fig 27). Ionized gas is perfectly correlated with molecular gas, and can be considered as the interface between molecular filaments and the very hot ambient environment.

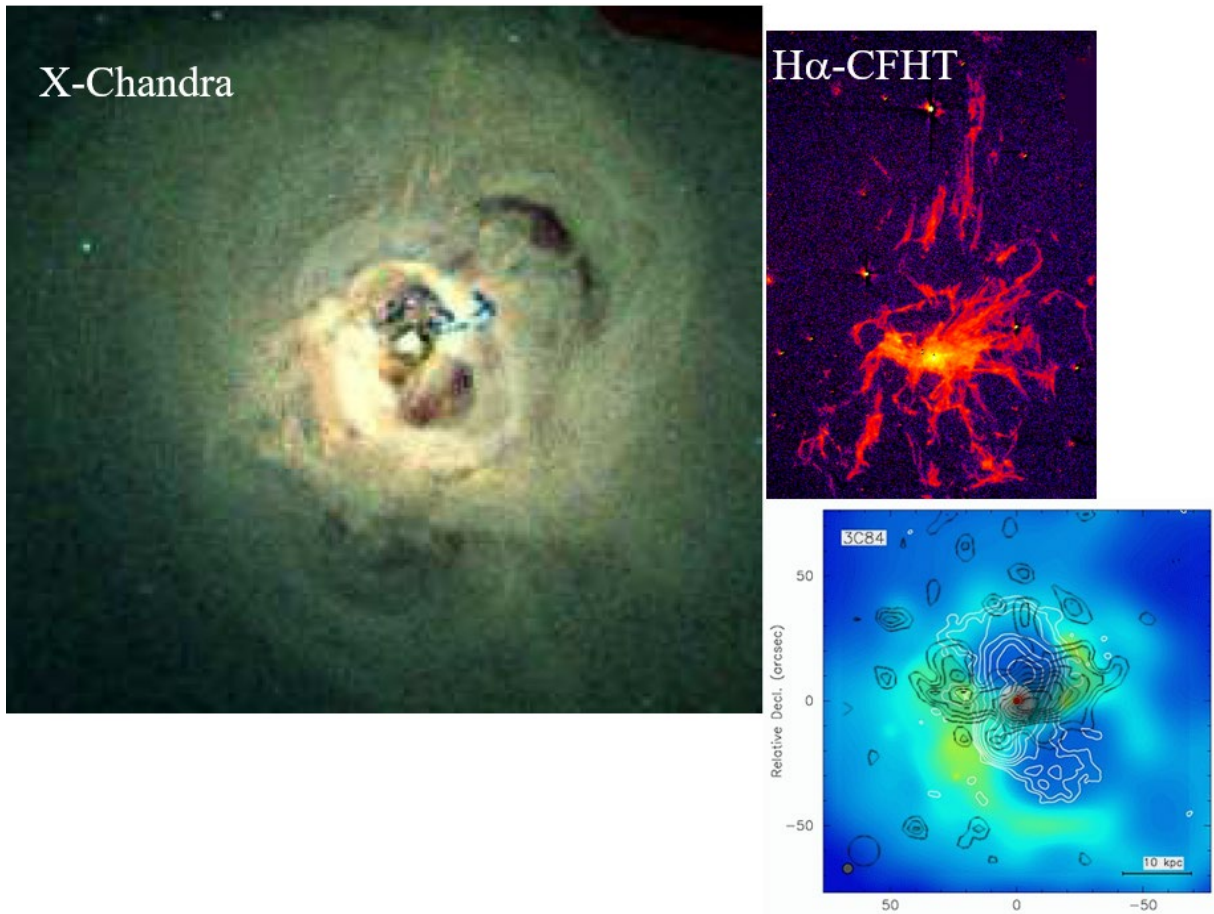

**Figure 27:** The cooling flow in the Perseus cluster. Left: the image obtained in X-rays by the Chandra satellite (Fabian et al 2003). The cavities dug into the hot gas are spectacular. Top right: at the same scale, the Hα image obtained with the CFHT telescope by Gendron-Marsolais et al (2018). Bottom right: on the image X-rays in blue and yellow, the contours of the radio continuum of the VLA (radio jet) are superimposed in white and those of the molecular gas are in black (line of the CO molecule (2-1), IRAM-30m, Salomé et al 2006). The radio jet fills the cavity, as the cold molecular gas concentrates on the edges of the cavities, falling towards the center.

From the energy balance perspective, the AGN feedback model to moderate the cooling flow is quite compatible with the orders of magnitude provided by observations. It can be seen that the radiative losses of gas in the cooling zones are compensated by thermal energy plus the mechanical energy of cavity expansion (PV). The cavities are emptied of hot gas, the X-ray emission comes only from the shell surrounding the cavities. After de-projection, the surface brightness on the cavities is compatible with the emission from the wall alone. The mass of the shells is comparable to the mass ejected from the cavities. The X-eay gas was pushed out of the cavities by the plasma ejected by the radio source and compressed into the shells (see McNamara et al 2016).

Another phenomenon is to be taken into account in the cooling of the gas moderated by the feedback of the central AGN. Clusters are frequently disturbed by the accretion of a group, or the merger with another cluster. In the relaxation that follows, the off-center BCG can oscillate around the center, with a period of approximately 300 million years, comparable to the gas cooling time. Thus, it can form a cooling trail behind the BCG, very visible in Hα, as shown in the case of Abell 1795 in Figure 28. Once again, the molecular gas is very correlated with the ionized gas, and is formed around the cavities dug by the radio jet (Russell et al 2017). These motions generated by the interaction dynamics, contribute to turbulence and inhomogeneities, which promote cooling far from the center of the BCG.

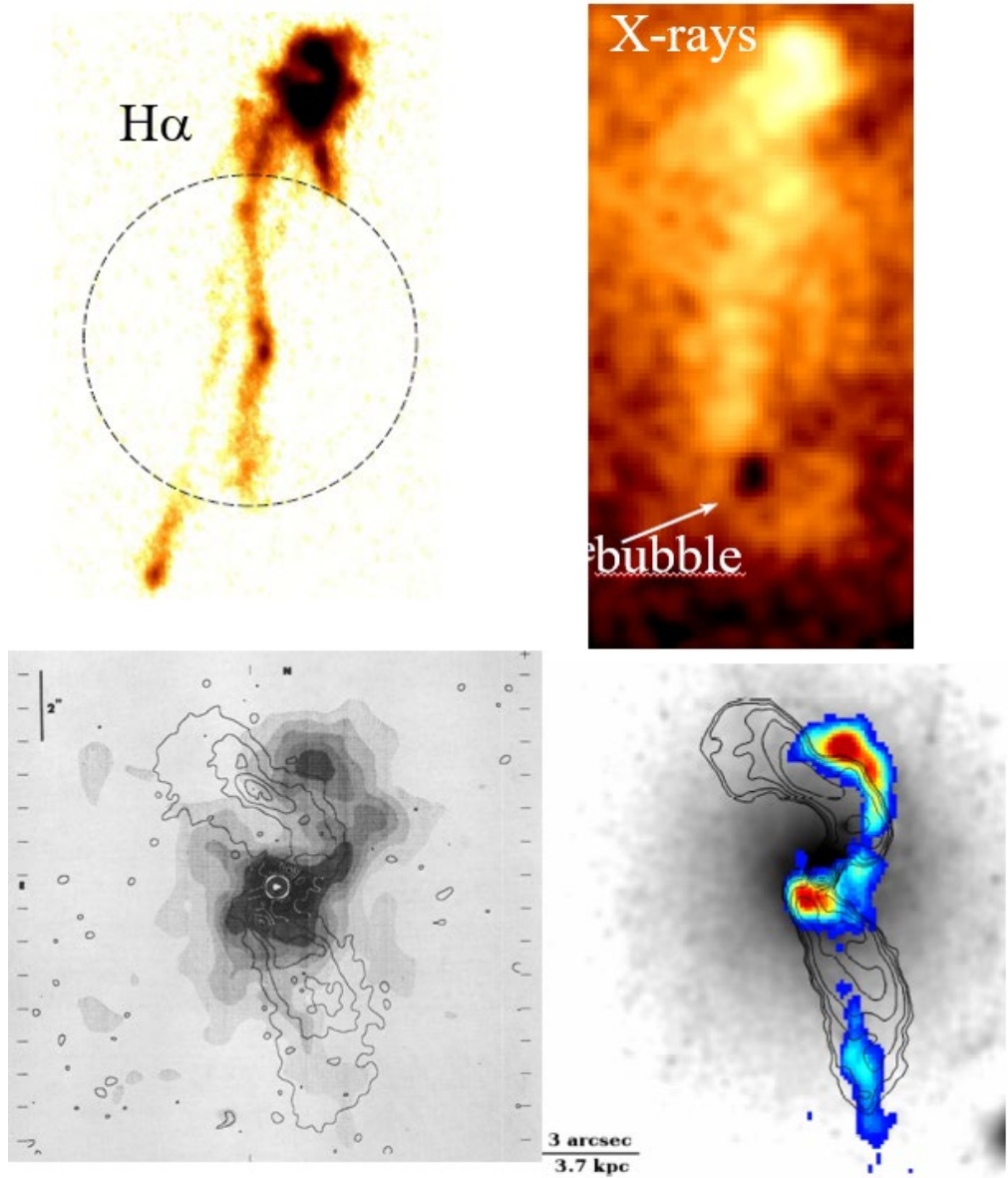

**Figure 28:** The cooling trail cluster, Abell 1795. Top left: the BCG is not in the center of the cluster, symbolized by the dashed circle, with radius of 22 kpc, but in the north, 33 kpc from the center (Cowie et al 1983). Top right, the trail is also very visible in X-rays with the Chandra satellite (same scale). Bottom left: zoomed in image of the central part of the galaxy, the contours of the VLA continuum are superimposed on the Hα image in gray (van Breugel et al 1984). Bottom right, color image at the same scale of the CO(2-1) map obtained with ALMA, superimposed on a grayed HST image (Russell et al 2017).

## 4.3 Molecular outflows

In the case of AGNs at the center of clusters, it is clear that the feedback phenomenon is essential to moderate the cooling of the hot gas and to avoid the formation of too massive

- 35 -

central galaxies. Is the feedback from low-luminosity AGNs at the center of nearby spiral galaxies (Seyfert or LINER) coupled as well to the host galaxy and as efficient? In section 2, we showed how statistically molecular outflows are correlated with the AGN luminosity, and that often the radio mode was most effective at low luminosities ($L < L_{Edd}$), provided that the radio jet is partly oriented in the plane of the host galaxy, and can sweep and drag the molecular gas.

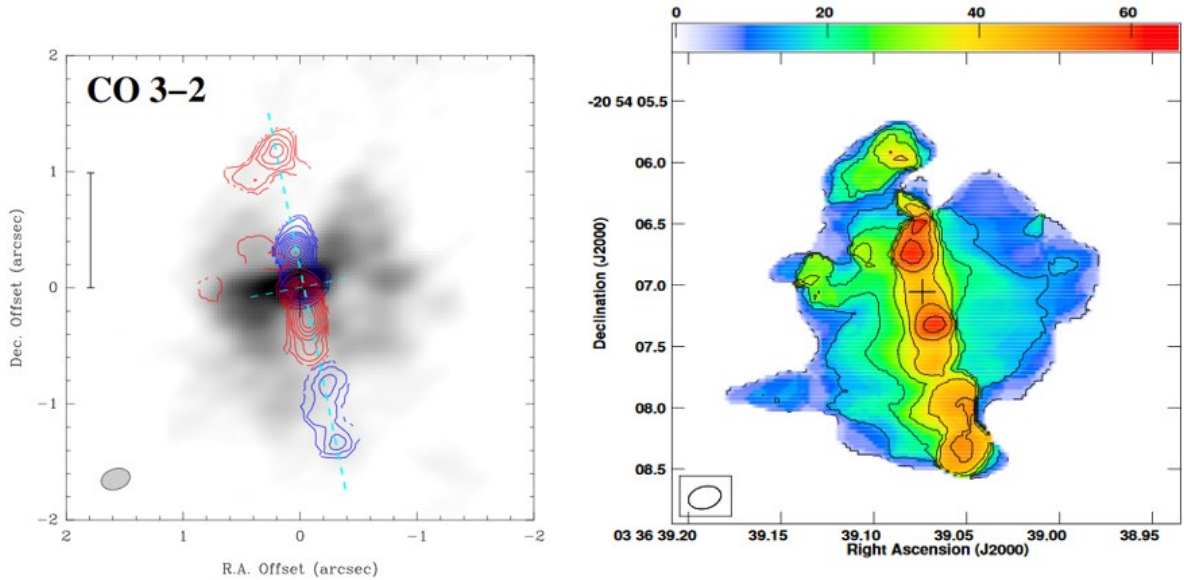

**Figure 29:** NGC 1377 seen by ALMA. Left: the integrated CO(3-2) emission for velocities close to the systemic velocity ($\pm$30km/s) is in gray. The emission in the approaching velocities is indicated by the contours in blue (-80 to -150km/s), and receding velocities in red (80 to 150km/s). The vertical bar indicates a scale of 100pc. The ellipse at the bottom gives the angular resolution. Right: Color map of the velocity dispersion (see palette) contours range from 4.4 to 57 km/s. The velocity dispersion underlines the molecular jet. According to Aalto et al (2016).

This is indeed the case with a large number of nearby galaxies, such as NGC 1377, illustrated in Figure 29. This galaxy is very special, because its centimetric radio emission is not detected. However, there is a very collimated molecular outflow, which cannot come only from star formation and the feedback of supernovae. The molecular jet even reverses: it approaches, then moves away from the observer, on either side of the disk (with alternating velocities shifted towards blue and then towards red). This reversal indicates a precession, with a jet moving in a plane close to the plane of the sky. This precession, which could be at an angle of approximately 10°, is not compatible with star formation feedback, but demonstrates the existence of a radio jet, below the detection threshold. The galaxy NGC 1068, Seyfert 2 prototype due to obscuration, also has a jet in the plane of the galaxy (e.g., Garcia-Burillo et al 2014). Either way, the AGN feedback is effective in delaying star formation, by heating and displacing the available gas. However, the displaced quantities are quite small (a few percent at most) compared to the total gas content of the galaxy, and furthermore the velocities are far from the escape velocity, so the gas will become available again to form other stars.

For nearby galaxies, it seems difficult to demonstrate that AGN feedback is efficient in stopping star formation. On the contrary, there are many statistics to show an excellent correlation between starburst formation and nucleus activity, and perhaps only for strong luminosities, the molecular gas content would be more fragile (e.g., Fiore et al 2017). Perhaps it could be a matter of timescales, the activity of the nuclei could be shifted compared to a starburst coming from a galaxy merger, but the common period is still long, up to 1.5 billion years (Volonteri et al 2015). Nearby AGN host galaxies have quite normal star-forming properties even in their central regions (Casasola et al 2015). In numerical simulations, depending on the algorithms and resolutions used, the feedback from AGNs can be very efficient (Nayakshin 2014), or not efficient at all (Roos et al 2015), sometimes the feedback is even positive (Gaibler et al, 2012, Bieri et al 2015).

## 5 Conclusion

The relationship between masses of black holes and host spheroids is now well established, and interpreted as the co-evolution and symbiosis between black holes and their galaxies. The exceptions to this relationship are very indicative of the mechanisms behind this symbiosis. The simultaneous growth of the bulge and the black holes comes naturally from the gas accretion which feed both, but also from the mergers of galaxies. Limiting and moderating star formation by AGN feedback has often been proposed to explain the relationship between the masses.

The feedback mechanisms can be broken down into two categories: (1) the quasar, or radiative mode, by relativistic winds, coming from the brightest AGNs; (2) the radio, or mechanical, mode exerted by radio jets, coming from AGNs of low luminosities relative to their Eddington luminosity.

Molecular outflows are very frequently observed around nearby AGNs, with velocities ranging from 200 to 1200km/s in the most extreme cases, typically ejecting $10^8$ M$_\odot$ of gas. The mass ratio between the rate of ejected gas and the rate of star formation can range from 1 to 5. Statistically, the outflows conserve energy, and their angular momentum is increased compared to that of the wind, by a factor $\sim$ 20. However, the feedback from AGNs, if it allows a little delay in star formation, does not seem efficient in stopping it completely.

The radio mode is very efficient in clusters to moderate the cooling of hot gas in the vicinity of the central massive galaxy. This moderation is produced by the action of radio jets, whose non-thermal plasma sculpts cavities into the hot gas in the cluster. The energy of the AGN heats up the gas, but the bubbles which rise by buoyancy also generate the cooling of the gas, which forms molecular filaments that fall back on the black hole to feed it, thus closing the feedback cycle.